\documentclass[prb,aps,twocolumn]{revtex4-2}

\usepackage[utf8]{inputenc}

\usepackage{mathtools}
\usepackage{amsmath}
\usepackage{amssymb}
\usepackage{bm}
\usepackage{nicefrac}
\usepackage{cancel}

\usepackage{adjustbox}
\usepackage{makecell}
\usepackage{multirow}
\usepackage{ragged2e}

\usepackage{color}
\usepackage{xcolor}
\usepackage[normalem]{ulem}

\usepackage{orcidlink}

\usepackage{graphicx}
\usepackage{tikz}
\usetikzlibrary{arrows.meta,calc}
\usepackage{hyperref}
\hypersetup{
  colorlinks=true,
  linkcolor=blue,
  filecolor=magenta,
  urlcolor=cyan,
  citecolor=blue,
}
\DeclareUnicodeCharacter{2009}{\,}

\newcommand{\be}{
\begin{equation}}
  \newcommand{\ee}{
\end{equation}}
\newcommand{\ba}{
\begin{eqnarray}}
  \newcommand{\ea}{
\end{eqnarray}}
\newcommand{\beq}{
\begin{eqnarray}}
  \newcommand{\eeq}{
\end{eqnarray}}

\newcommand{\mcusp}{\mathcal{M}}
\newcommand{\LLL}{\text{P}_{\text{LLL}}}

\newcommand{\lB}{\ell_{\mathrm{B}}}

\newcommand{\abs}[1]{\vert #1 \vert}
\newcommand{\bra}[1]{\langle #1 \vert}

\newcommand{\conj}[1]{\bar{#1}}

\newcommand{\ket}[1]{\vert #1 \rangle}

\newcommand{\figurefolder}{./figures/}

\begin{document}
\title{Dressing composite fermions with artificial intelligence}
\author{Mytraya Gattu~\orcidlink{0000-0001-6994-389X}}
\affiliation{Department of Physics, 104 Davey Lab, Pennsylvania State University, University Park, Pennsylvania 16802, USA}
\date{\today}
\begin{abstract}
 Recent variational studies have demonstrated that the strongly correlated ground states of the fractional quantum Hall (FQH) effect can be captured using machine learning approaches starting from no prior knowledge of the underlying physics. We introduce a complementary framework that instead starts from Jain's composite-fermion (CF) wavefunctions, which accurately describe FQH states as weakly interacting states of CFs at fillings $\nu = n/(2pn+1)$ in an idealized limit. As we move away from this idealized limit to one more in line with experimental reality, we expect CFs to become dressed much like the electrons of a noninteracting system, which are dressed by neutral excitations as interaction is turned on adiabatically, as in Landau's Fermi-liquid theory. We model this dressing using a Feynman-Cohen-style backflow approach, implemented through symmetry-preserving neural networks—a framework we refer to as CF-Flow. CF-Flow achieves competitive accuracy with substantially greater computational efficiency and scales to systems of $\gtrsim 26$ electrons. At fillings $\nu = 1/3$ and $2/5$, as a function of Landau-level mixing strength, CF-Flow produces ground-state energies with low local-energy variance that are nearly indistinguishable from those obtained using the fixed-phase diffusion Monte Carlo (fp-DMC) method, even though the latter constrains the wavefunction phase to that of the lowest Landau level—thereby providing insight into why fp-DMC has been successful in giving an accurate quantitative account of several experiments. Finally, the symmetry-preserving architecture of CF-Flow enables access to excited states and computation of the transport gap at $\nu = 1/3$, where we find, unexpectedly, that it decays exponentially toward a finite value in the limit of large Landau-level mixing, suggesting a first-order transition from the FQH liquid to a non-FQH state.
\end{abstract}
\maketitle

\section{Introduction}\label{sec:introduction}

That any progress has been achieved in understanding the behavior of interacting systems of quantum mechanical particles—the subject of much of modern condensed matter physics—is extremely remarkable. At the outset, the problem seems impossible: given that even the problem of three interacting electrons eludes an exact solution, how are we to understand the behavior of $10^{21}$ electrons?

For metals, Landau's Fermi-liquid theory offers valuable insight~\cite{Baym91}. The theory begins with noninteracting electrons, for which we know the full solution: the ground state is a Fermi sea, and the excitations are electron-hole pairs. One then assumes that as the interaction is slowly turned on, no phase transition occurs, and the electrons of the noninteracting system evolve into Landau quasiparticles—entities that retain the same quantum numbers as electrons but are ``dressed'' by a screening cloud of particle-hole excitations emerging from the Fermi sea. The properties of these quasiparticles near the Fermi energy can then be estimated through perturbative techniques.

In this paper we consider a strongly interacting system, namely electrons in two dimensions exposed to a strong perpendicular magnetic field-the setup of the fractional quantum Hall effect (FQHE)~\cite{Stormer83}. This is actually a much more difficult problem than that in the preceding paragraph. The reason is that when electrons occupy the lowest quantum of the kinetic energy (namely the lowest Landau level (LLL)), their behavior is determined entirely by interactions, and it is not adiabatically connected to noninteracting electrons (which form a macroscopically degenerate system). We do not have a natural starting point akin to the Fermi sea of the Landau Fermi-liquid theory. How, then, are we to tackle this problem?

Several recent studies~\cite{Teng25, Qian25} have applied neural quantum states (NQS)~\cite{Carleo17, Carleo19, Lange24} to this problem, leveraging the universal expressivity of neural networks to represent quantum many-body wavefunctions within variational Monte Carlo (VMC). These works employ the Psiformer architecture~\cite{Pfau23}, a transformer-based NQS that uses attention mechanisms~\cite{Vaswani17} to model electron correlations. Remarkably, they demonstrate that the dramatic reorganization from a macroscopically degenerate noninteracting system to gapped, incompressible FQH states can be captured by starting from bare electrons and dressing them using neural networks. As noted in Ref.~\cite{Qian25}, however, learning all correlations ab initio can become computationally demanding and currently limits accessible system sizes, though recent pre-training strategies~\cite{Nazaryan25} offer promising avenues for scaling Psiformer-like Ans\"atze to larger systems.

We take an alternative approach, building on the fact that we do know the solution with a high degree of accuracy for an idealized model: electrons in a strictly two-dimensional space with pure Coulomb interactions and no mixing between different LLs. The accuracy of the Jain wavefunctions~\cite{Jain89, Jain07} has been confirmed by comparisons with exact solutions known for small systems from computer calculations~\cite{Fano86, Dev92, Wu93, Jain07, Balram24}. Remarkably, this solution is formulated in terms of weakly interacting (or, in the idealized limit, noninteracting) composite fermions (CFs), related to the solution of noninteracting electrons at an effective magnetic field. As the system deviates from this ideal limit—whether through finite quantum well width~\cite{Suen92,Suen94,Luhman08,Shabani13,Shabani09,Shabani09B,Sharma24}, LL mixing~\cite{Zhao18, Zhao23}, or considering CF states in higher Landau levels such as the second LL~\cite{Toke05}—the interactions between CFs increase and they become dressed by CF particle-hole excitations. When the system is close to the ideal limit, we expect only quantitative corrections, but eventually these effects will cause a transition into a different state. We model this dressing through a backflow transformation that captures how each CF's motion influences others—an approach pioneered by Feynman and Cohen for liquid helium~\cite{Feynman56}. Our framework, which we call CF-Flow, combines Jain's CF wavefunctions with symmetry-preserving neural networks~\cite{Zaheer17, Bronstein17, Villa21, Debenedetti25} to implement this backflow~\cite{Feynman56, Lopez06, Luo19, Pescia24}. Importantly, CF-Flow is a general approach applicable whenever interactions between CFs go beyond the perturbative regime where analytical techniques remain tractable~\cite{Sodemann13,Macdonald84,Melik-Alaverdian95,Murthy02,Bishara09,Wojs10,Simon13,Peterson13,Peterson14}.

In this work, we focus on LL mixing as a first application of CF-Flow and, for simplicity, do not include the effect of finite quantum well width. LL mixing is quantified by the dimensionless parameter $\kappa$, 
\begin{equation}\label{eq:kappa-definition}
\kappa = \frac{e^{2}}{\epsilon \lB \hbar \omega_{c}},
\end{equation}
where $\lB=\sqrt{\hbar c/(eB)}$ is the magnetic length, $\hbar \omega_{c}=eB/cm^{\ast}$ is the cyclotron frequency, and $m^{\ast}$ is the electron effective mass. The ``LLL limit'' corresponds to $\kappa \to 0$, but experiments operate far from this idealized regime. Even at the highest accessible magnetic fields, electron-type GaAs samples (which have very small $m^{\ast}$) reach only $\kappa\gtrsim 0.4$, while hole systems reach $\kappa\gtrsim 2.5$; typical experiments operate in the range $\kappa\sim 0.50-10$~\cite{Jain07}. It has long been thought that this mismatch between $\kappa$ values in theoretical calculations and experimental conditions contributes to differences in calculated and measured energy gaps~\cite{Morf02, Melik-Alaverdian95, Murthy02}. 

Moreover, LL mixing is known to induce phase transitions—such as the transition at filling factor $\nu = 1/4$ from a CF metal~\cite{Halperin93,Willett93,Hossain19,Halperin20} to a paired (superconducting) state of CFs (manifesting as an even-denominator FQH state)~\cite{Zhao23, Wang23}, and transitions at low fillings between liquid FQH states and crystalline phases~\cite{Zhao18, Rosales21B, Santos92}. Perhaps most notably, LL mixing is thought to underlie the long-standing discrepancy between theory and experiment regarding the nature of the $\nu = 5/2$ ground state~\cite{Willett87, Pan99, Morf98, Moore91, Moller08, Peterson08A, Wojs10, Rezayi11, Pakrouski15, Rezayi17}—a proposed platform for topological quantum computing~\cite{Dassarma05}. However, these transitions and typical experimental conditions occur at values of $\kappa$ beyond the perturbative regime for which analytical techniques have been developed, motivating the use of nonperturbative variational approaches, such as CF-Flow, to access this physics. (The filling factor $\nu$ is defined as $\rho \phi_{0}/B$, where $\rho$ is the transverse electron density and $\phi_{0}=hc/e$ is the magnetic flux quantum.)

Previous approaches have employed the fixed-phase diffusion Monte Carlo (fp-DMC) method~\cite{Ortiz93, Kolorenc11, Melik-Alaverdian01,Melik-Alaverdian97,Zhao18, Zhang16, Zhao23}, which gives the lowest energy in a given phase sector. While this method does give a strict upper bound on the ground state energy, its accuracy depends on the choice of the phase of the wavefunction (with any choice being somewhat ad hoc). CF-Flow provides an independent approach to this problem, using an order-of-magnitude fewer parameters than other ML architectures while training substantially faster and scaling to $26$ or more electrons. At the Jain filling fractions, CF-Flow yields accurate ground state energies as a function of the strength of LL mixing, ascertained from the low local-energy variance.

While our primary objective in this article is to develop CF-Flow, the framework already yields useful insight into LL mixing. A notable observation arises from comparison with fp-DMC~\cite{Zhao18}: despite fp-DMC constraining the wavefunction phase to the LLL, the two approaches produce essentially identical ground-state energies at the Jain fillings. This agreement supports the central assumption underlying fp-DMC and helps explain its success in reproducing experimental measurements—from spin-polarization boundaries~\cite{Zhang16} to transitions between CF metals, CF superconductors~\cite{Zhao23}, and liquid-crystal phases~\cite{Zhao18,Ma20B,Rosales21B}—even though real systems operate far from the LLL limit.

CF-Flow's symmetry-preserving architecture also enables access to excited states by initializing from the appropriate CF wavefunction in a chosen angular-momentum sector. We illustrate this by computing the transport gap at $\nu = 1/3$ across a wide range of LL-mixing strengths. The gap decreases exponentially with increasing $\kappa$ but approaches a finite value rather than closing—consistent with a first order transition from the topological FQH liquid to a different phase. This prediction can be tested through activated transport measurements as LL mixing is varied.

The rest of the paper is organized as follows. In Sec.~\ref{sec:background}, we review Haldane's spherical geometry~\cite{Haldane83}, and in Sec.~\ref{sec:cf-theory} we summarize CF theory and the construction of CF wavefunctions for ground and excited states at the Jain fillings. Sec.~\ref{sec:backflow} adapts the Feynman-Cohen backflow approach to the sphere, which Sec.~\ref{sec:cf-flow} then combines with CF wavefunctions through symmetry-preserving neural networks to define CF-Flow. Sec.~\ref{sec:results} presents our training methodology and results: ground-state energies at $\nu=1/3$ and $2/5$ as functions of $\kappa$, and the transport gap at $\nu=1/3$. Finally, Sec.~\ref{sec:discussion} offers discussion and conclusion.

\begin{figure*}
  \includegraphics[width=0.8\linewidth]{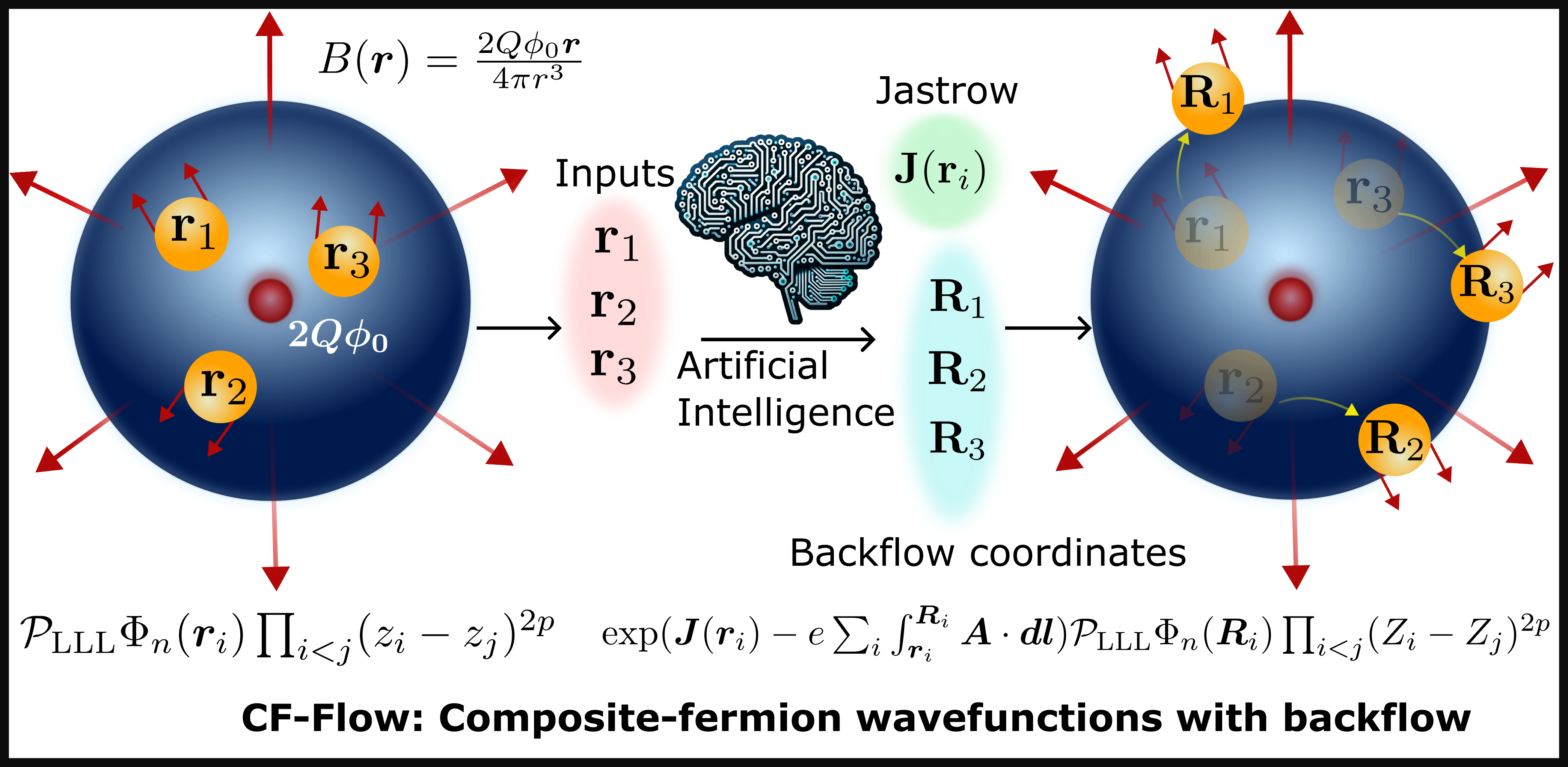}
  \caption{Schematic illustration of CF-Flow. In the Feynman--Cohen backflow picture~\cite{Feynman56}, turning on interactions causes particles to shift from their noninteracting positions $\bm{r}_{i}$ to backflow positions $\bm{R}_{i}$. CF-Flow implements this idea by combining the backflow approach with composite-fermion (CF) wavefunctions defined on the Haldane sphere using symmetry-preserving machine-learning methods, enabling us to study regimes in which interactions between CFs become significant.}
  \label{fig:sphere-schematic}
\end{figure*}

\section{Background}\label{sec:background}
In this section, we introduce the spherical geometry~\cite{Haldane83} used to study the bulk properties of fractional quantum Hall (FQH) liquids and describe the corresponding single-particle wavefunctions. We then define the many-body problem and briefly discuss the commonly employed lowest Landau level (LLL) limit.

\subsection{Single-particle physics on the Haldane sphere}\label{sec:single-particle}

Consider the setup of the FQHE: interacting electrons confined to two dimensions in the presence of a strong perpendicular magnetic field. To study such a system, we employ Haldane's spherical geometry~\cite{Haldane83}, where electrons are restricted to the surface of a sphere of radius $R$ with a magnetic monopole of charge $Q \in \mathbb{Z}/2$ at its center, producing a uniform radial magnetic field (see Fig.~\ref{fig:sphere-schematic} for an illustration of the Haldane sphere).

For studying the bulk properties of FQH liquids, this geometry offers several advantages. The compact, edge-free nature of the sphere suppresses finite-size corrections from bulk-edge coupling (in contrast to the disk or planar geometries). Its genus-zero topology~\cite{Wen98} ensures a unique gapped ground state for FQH liquids, unlike the torus, where FQH states at filling factor $\nu = p/q$ form a $q$-fold degenerate ground-state manifold. Moreover, different topological states at a given filling $\nu$ can be distinguished by their ``Wen-Zee shift'' $\mathcal{S}$~\cite{Wen92}: an FQH state of $N$ electrons occurs at flux $N_\phi = N/\nu - \mathcal{S}$. For these reasons, we adopt the spherical geometry in this work. We describe below the single-particle physics.

The magnetic monopole generates a uniform radial magnetic field of strength $B = 2Q\phi_{0}/4\pi R^{2}$, where $\phi_{0} = \hbar c / e$ is the magnetic flux quantum. In the commonly used gauge $\bm{A} = -(\hbar c Q \cot\theta / eR)\,\hat{\phi}$~\cite{Haldane83, Jain07}, the kinetic Hamiltonian $\mathcal{K}$ is
\begin{equation}\label{eq:kinetic-operator}
  \mathcal{K} = \frac{\hbar \omega_{c}}{2Q}\left[-\frac{1}{\sin\theta}\frac{\partial}{\partial\theta}\sin\theta\frac{\partial}{\partial\theta}
  + \left(Q\cot\theta + \frac{i}{\sin\theta}\frac{\partial}{\partial\phi}\right)^{2}\right],
\end{equation}
where $\omega_{c} = eB/m_{e}c$ is the cyclotron frequency.

The eigenstates of $\mathcal{K}$ are the monopole harmonics $Y_{Q,l,m}(\theta,\phi)$~\cite{Wu76, Haldane83, Jain07}. These represent the single-particle quantum states available to an electron on the sphere—the spherical analogue of Landau orbitals in planar geometry. They are labeled by angular momentum quantum number $l = |Q|, |Q| + 1, \dots$ and azimuthal quantum number $m = -l, -l + 1, \dots, l$, and have kinetic energy:
\begin{equation}\label{eq:kinetic-energies}
  \mathcal{K}Y_{Q,l,m}(\theta,\phi)
  = \hbar \omega_{c}\frac{l(l+1) - Q^{2}}{2Q}\,Y_{Q,l,m}(\theta,\phi).
\end{equation}
States with the same $l$ form a $(2l+1)$-fold degenerate shell (corresponding to the $2l+1$ possible values of $m$), and constitute the $(l-\abs{Q})^{\rm th}$ Landau level in spherical geometry.

In the lowest Landau level, the monopole harmonics take a particularly simple form in terms of the spinor variables $u=\cos(\theta/2)e^{i\phi/2}$ and $v=\sin(\theta/2)e^{-i\phi/2}$:
\begin{equation}\label{eq:lll-harmonics}
  Y_{Q,Q,m}(\theta,\phi)
  = (-1)^{Q-m}\sqrt{\frac{2Q+1}{4\pi}}u^{Q+m}v^{Q-m}.
\end{equation}
In higher Landau levels, they are multi-term polynomials involving $u,v,u^\ast,v^\ast$~\cite{Jain07}.

Alternatively, they can be viewed mathematically as a section of the Wigner $D$-matrices~\cite{Wu76, Wu77, Boyle16, Gattu25}:
\begin{equation}\label{eq:monopole-harmonics-wigner-D-relation}
  \begin{split}
    Y_{Q,l,m}(\theta,\phi)
    &= \sqrt{\frac{2l+1}{4\pi}}\,D^{l}_{-m,-Q}(\phi,\theta,0), \\
    D^{l}_{m_{1},m_{2}}(\alpha,\beta,\gamma)
    &= \langle l,m_{1}|\,
    e^{-i\alpha J_{z}} e^{-i\beta J_{y}} e^{-i\gamma J_{z}}
    \,|l,m_{2}\rangle,
  \end{split}
\end{equation}
where $J_{x}$, $J_{y}$, and $J_{z}$ are the generators of rotations satisfying $[J_{i},J_{j}] = i\epsilon_{ijk}J_{k}$, and $\ket{l,m}$ are eigenstates with angular momentum $l$ and azimuthal quantum number $m = -l, \dots, l$~\cite{Sakurai20}.

\subsubsection*{A key distinction from planar Landau levels}

An important mathematical feature distinguishes monopole harmonics from their planar counterparts. Due to the singular nature of the vector potential $\bm{A}$ at the poles (i.e., the Dirac strings), an arbitrary function $f(\theta, \phi)$ that contains components in a monopole sector $Q' \neq Q$ generically possesses infinite kinetic energy—unless it vanishes at both poles~\cite{Qian25}. 

To see this explicitly, consider evaluating the kinetic energy of $Y_{Q', l', m'}(\theta, \phi)$ with respect to the kinetic operator at monopole strength $Q$ [Eq.~\eqref{eq:kinetic-operator}]. For clarity, we adjust the sphere radius such that electrons always experience the same magnetic field strength $B$, and use primes to denote operators and wavefunctions at monopole strength $Q'$. From Eqs.~\eqref{eq:kinetic-operator} and~\eqref{eq:kinetic-energies}, we obtain:
\begin{align*}
&\mathcal{K}Y_{Q', l', m'} \\
&= \frac{\hbar \omega_{c}}{2Q}\left[-\nabla^{2}+Q^{2}\cot^{2} \theta -\frac{2Qm' \cot \theta}{\sin \theta}\right]Y_{Q', l', m'}.
\end{align*}
Since $\mathcal{K}'Y_{Q', l', m'} = \frac{\hbar \omega_{c}[l'(l'+1)-Q'^{2}]}{2 Q'}Y_{Q', l', m'}$, where
\begin{align*}
\mathcal{K}' = \frac{\hbar \omega_{c}}{2Q'}\left[-\nabla^{2}+Q'^{2}\cot^{2} \theta -\frac{2Q' m' \cot \theta}{\sin \theta}\right],
\end{align*}
we can rearrange to find:
\begin{align*}
&\mathcal{K}Y_{Q', l', m'} = \frac{\hbar \omega_{c}[l'(l'+1)-Q'^{2}]}{2Q}Y_{Q', l', m'} \\
& + \frac{\hbar \omega_{c}(Q-Q')\cos \theta}{2Q \sin^{2} \theta}\left[(Q+Q')\cos\theta-2m'\right]Y_{Q', l', m'}.
\end{align*}

Using the relation in Eq.~\eqref{eq:monopole-harmonics-wigner-D-relation}, the expectation value of the kinetic energy becomes:
\begin{align*}
&\langle Y_{Q', l', m'}|\mathcal{K}|Y_{Q', l', m'}\rangle = \frac{\hbar \omega_{c}[l'(l'+1)-Q'^{2}]}{2 Q} \\
&\quad + \frac{\pi \hbar \omega_{c}(Q-Q')}{Q} \\
&\times \int_{0}^{\pi}d\theta \frac{\cos \theta}{\sin \theta}\left[(Q+Q')\cos\theta-2m'\right] [d^{l'}_{-m',-Q'}(\theta)]^{2}.
\end{align*}

When $|m'|=|Q'|$ and $Q' \neq Q$, the integral diverges due to the $\cot \theta$ singularity at one of the poles. Since $Y_{Q', l', \pm Q'}$ (proportional to $d^{l'}_{-m',-Q'}(\theta)$) are the only monopole harmonics nonzero at the poles, any function $f(\theta, \phi)$ with support at the poles (i.e., that is non-zero at the poles) generically exhibits divergent kinetic energy unless it lies entirely within the correct monopole sector $Q$.

This imposes a constraint on variational wavefunctions in spherical geometry: they must, by construction, lie entirely within the correct monopole sector to avoid formally infinite kinetic energy. While energy minimization generically drives wavefunctions toward the physical sector, intermediate configurations during optimization may develop pathological spikes i.e., have anomalously large energies (see Ref.~\cite{Qian25} for discussion in the context of Psiformer-like ans\"atze). For CF-Flow, which builds upon backflow transformations of Jain's CF wavefunctions (which are of course within the correct monopole sector), this requirement is automatically satisfied (see Sec.~\ref{sec:backflow}).

Having established the single-particle physics, we now turn to the many-body problem.

\subsection{The many-body problem}\label{sec:many-body-problem}
In this work, our focus is on understanding how the Jain FQH states at fillings $\nu = n/(2pn+1)$ (which have a Wen-Zee shift $\mathcal{S}=n+2p$) evolve as we move away from the zero LL mixing limit. We restrict our attention to the ideal 2D limit, in which electrons interact via the Coulomb interaction.

Formally, the problem is to identify the ground (and excited) state(s) of the Hamiltonian (in units of $e^{2}/\varepsilon \lB$) for $N$ electrons facing a monopole of charge $2Q=N/\nu-(n+2p)$:
\begin{equation}\label{eq:many-body-H-kappa}
  \mathcal{H} = \sum_{i=1}^{N} \frac{\mathcal{K}_{i}}{\kappa}+\sum_{i<j=1}^{N}\frac{1}{\sqrt{Q} |\Omega_{i}-\Omega_{j}|}.
\end{equation}
Here, $\kappa$ is the Landau-level mixing parameter, defined as the ratio of the Coulomb energy scale $e^{2}/\varepsilon \lB$ to the cyclotron energy $\hbar \omega_c$. We use the shorthand notation $\Omega_{i} \equiv (\theta_{i}, \phi_{i})$ to refer to the $i^{\rm th}$ electron's position. The Hamiltonian is rotationally invariant (up to a redefinition of the gauge defining the vector potential). Consequently, the many-body eigenstates are labeled by total angular momentum $L$ and azimuthal quantum number $M$. For example, the uniform FQH ground states occur at $L=M=0$.

While experiments are typically conducted in the $\kappa \sim 0.50-10$ regime, most of our theoretical understanding of the observed phenomenology of FQH liquids has been achieved by working in the ``LLL limit'' corresponding to $\kappa = 0$. In this limit, the kinetic energy is fully quenched (i.e. all electrons inhabit only the LLL) and the system is governed solely by Coulomb interactions projected into the LLL:
\begin{equation}\label{eq:lll-hamiltonian}
  \mathcal{H}_{\rm LLL} = \LLL\left(\sum_{i<j=1}^{N}\frac{1}{\sqrt{Q} |\Omega_{i}-\Omega_{j}|}\right)\LLL,
\end{equation}
where $\LLL$ is the operator that projects each electron's state into the LLL. The zero-point energy has been absorbed into the Hamiltonian definition (we follow this convention henceforth).

The LLL limit affords several advantages. First, the LLL limit drastically reduces the available degrees of freedom, making the Fock space finite-dimensional. This has allowed exact diagonalization (ED) to emerge as a powerful tool for studying the FQHE: both as an independent method to probe the physics and as a rigorous way to test trial wavefunctions such as those of Laughlin~\cite{Laughlin83}, Jain~\cite{Jain89} and Moore-Read~\cite{Moore91}. Such trial wavefunctions form the foundation of much of our present theoretical understanding.

Second, the LLL limit provides clean theoretical insight: to the extent that experimental systems can be connected adiabatically to this limit, the emergence of gapped FQH phases demonstrates that these states are fundamentally nonperturbative, arising entirely from electron-electron interactions. This includes the emergence of composite fermions (CFs) and their $\Lambda$ levels—Landau-level-like structures inhabited by CFs, formed solely through interactions (discussed further in Sec.~\ref{sec:cf-theory}).

Third, while the qualitative success of CF theory~\cite{Jain89, Jain07}—which relates the FQHE of electrons to the IQHE of CFs—does not strictly require the LLL limit, its quantitative accuracy does. LLL-projected CF wavefunctions not only exhibit overlaps exceeding $0.99$ with exact ground states for accessible system sizes~\cite{Fano86, Dev92, Wu93, Jain07, Balram24} but also reproduce a vast range of experimental observations—often quantitatively, and in other cases semi-quantitatively or qualitatively~\cite{Jain07}. The combination of this accuracy, physical transparency, and computational tractability has made the LLL limit the standard framework for theoretical investigations of the FQHE.

As an aside, we note that in the absence of LL mixing, even FQH states in higher Landau levels, such as the second Landau level (SLL), can be effectively described within the LLL framework by constructing an appropriate effective Hamiltonian that captures the modified interactions~\cite{Jain07,Yutushui25}.

Despite the success of the LLL framework, as discussed in Sec.~\ref{sec:introduction}, there are compelling reasons to move beyond it—motivating the development of nonperturbative variational methods such as CF-Flow.


We next review CF theory and the construction of CF wavefunctions, which provide the foundation for CF-Flow.
\section{Composite-fermion (CF) theory}\label{sec:cf-theory}

In this section, we provide a brief overview of CF theory and Jain's CF wavefunctions.

The CF theory~\cite{Jain89} posits that electrons constrained to the LLL capture an even number of vortices ($2p$, where $p \in \mathbb{Z}^{+}$) to transform into \emph{weakly interacting} composite fermions. This vortex attachment screens the external magnetic field, causing the CFs to experience a reduced monopole charge $Q^{\ast}=Q-p(N-1)$ and inhabit their own Landau-level-like structures, known as $\Lambda$ levels. These $\Lambda$ levels are separated by the CF cyclotron energy, $\hbar\omega_{c}^{\ast}$, which arises entirely from inter-electron interactions ($\propto e^{2}/\varepsilon \lB$) and scales as $Q^{\ast}$. The central dogma of the theory is that the integer quantum Hall effect (IQHE) of CFs inhabiting $\Lambda$ levels manifests as the FQHE of electrons. In particular, an integer filling ($\nu^{\ast}=n$) of $\Lambda$ levels creates an incompressible state of CFs, which manifests as the FQHE of electrons at the Jain fillings $\nu = n/(2pn+1)$. The resulting many-body state $\Psi_{\tfrac{n}{2pn+1}}$ is written as
\begin{widetext}
  \begin{equation}\label{eq:cf-ground-state}
    \Psi_{\tfrac{n}{2pn+1}}(\Omega_{1}, \dots, \Omega_{N}) = \LLL \underbrace{
      \begin{vmatrix}
        Y_{Q^{\ast}, Q^{\ast}, -Q^{\ast}}(\Omega_{1}) & \dots & Y_{Q^{\ast}, Q^{\ast}, -Q^{\ast}}(\Omega_{N}) \\
        \vdots &  \vdots &  \vdots \\
        Y_{Q^{\ast}, Q^{\ast}, Q^{\ast}}(\Omega_{1}) & \dots & Y_{Q^{\ast}, Q^{\ast}, Q^{\ast}}(\Omega_{N}) \\
        Y_{Q^{\ast}, Q^{\ast}+1, -(Q^{\ast}+1)}(\Omega_{1}) & \dots & Y_{Q^{\ast}, Q^{\ast}+1, -(Q^{\ast}+1)}(\Omega_{N}) \\
        \vdots & \vdots & \vdots \\
        Y_{Q^{\ast}, Q^{\ast}+n-1, (Q^{\ast}+n-1)}(\Omega_{1}) & \dots & Y_{Q^{\ast}, Q^{\ast}+n-1, (Q^{\ast}+n-1)}(\Omega_{N}) \\
    \end{vmatrix}}_{\displaystyle \phi_{n}(\Omega_{1}, \dots, \Omega_{N};Q^{\ast})}\underbrace{\prod_{i<j=1}^{N}(u_{i}v_{j}-u_{j}v_{i})^{2p}}_{\displaystyle \phi_{1}^{2p}}
  \end{equation}
\end{widetext}
Here, $\phi_{n}$ is a Slater determinant of $N$ electrons facing a monopole of charge $Q^{\ast} = N/2n-n/2$ and filling $n$ LLs, and $\phi_{1}^{2p}=\prod_{i<j}(u_iv_j-u_jv_i)^{2p}$ is the Jastrow factor which attaches $2p$ vortices to each electron and turns them into CFs. (Alternatively, a CF state can be thought of as a special case of the more general parton states~\cite{Jain89B, Balram18} and a product of one $\nu=n$ parton and of $2p$ $\nu=1$ partons.) $u_{i}=\cos(\theta_{i}/2)e^{\imath \phi_{i}/2}$, $v_{i}=\sin(\theta_{i}/2)e^{-\imath \phi_{i}/2}$ are the spinor representations of $\Omega_{i}$.

As mentioned previously, the LLL projection $\LLL$ is essential to the quantitative accuracy of CF wavefunctions. (The unprojected CF wavefunctions, though topologically equivalent to their projected counterparts, generally have higher energies~\cite{Jain07}.) In practice, this projection is implemented through the Jain-Kamilla (JK) approach~\cite{Jain97,Gattu25}, which yields a slightly modified but computationally tractable form of the wavefunction—enabling studies of systems with $\gtrsim 900$ CFs~\cite{Anakru25}. 

The key insight is to restructure the projection to avoid a many-body calculation (which scales as $N^{N}$). We begin by decomposing the Jastrow factor as:
\begin{align*}
  &\prod_{i<j=1}^{N}(u_{i}v_{j}-u_{j}v_{i})^{2p} = \prod_{i<j=1}^{N}(u_{i}v_{j}-u_{j}v_{i})^{2p-2}\\
  &\quad\times \prod_{i=1}^{N}\left(\underbrace{\prod_{j \neq i}(u_{i}v_{j}-u_{j}v_{i})}_{J_{i}}\right).
\end{align*}
Absorbing the ``individual'' Jastrow factors $J_{i}$ into each column of the Slater determinant, we can rearrange the CF wavefunction from Eq.~\eqref{eq:cf-ground-state} as:
\begin{widetext}
  \begin{equation}\label{eq:cf-ground-state-jain-kamilla-interim}
    \Psi_{\tfrac{n}{2pn+1}}(\Omega_{1}, \dots, \Omega_{N}) = \LLL {
      \begin{vmatrix}
        Y_{Q^{\ast}, Q^{\ast}, -Q^{\ast}}(\Omega_{1})J_{1} & \dots & Y_{Q^{\ast}, Q^{\ast}, -Q^{\ast}}(\Omega_{N})J_{N} \\
        \vdots &  \vdots &  \vdots \\
        Y_{Q^{\ast}, Q^{\ast}, Q^{\ast}}(\Omega_{1})J_{1} & \dots & Y_{Q^{\ast}, Q^{\ast}, Q^{\ast}}(\Omega_{N})J_{N} \\
        Y_{Q^{\ast}, Q^{\ast}+1, -(Q^{\ast}+1)}(\Omega_{1})J_{1} & \dots & Y_{Q^{\ast}, Q^{\ast}+1, -(Q^{\ast}+1)}(\Omega_{N})J_{N} \\
        \vdots & \vdots & \vdots \\
        Y_{Q^{\ast}, Q^{\ast}+n-1, (Q^{\ast}+n-1)}(\Omega_{1})J_{1} & \dots & Y_{Q^{\ast}, Q^{\ast}+n-1, (Q^{\ast}+n-1)}(\Omega_{N})J_{N} \\
    \end{vmatrix}}\phi_{1}^{2p-2}.
  \end{equation}
\end{widetext}
This rearrangement has a crucial property: in the $i^{\rm th}$ column, only the state of the $i^{\rm th}$ electron lies outside the LLL—all LLL wavefunctions are polynomials in only $u_{i}, v_{i}$ and not their complex conjugates [Eq.~\eqref{eq:lll-harmonics}]. Projecting a single electron's state into the LLL is exponentially simpler than projecting the full many-body state.

Jain and Kamilla therefore proposed moving the projection inside the determinant, projecting each column independently:
\begin{widetext}
  \begin{equation}\label{eq:cf-ground-state-jain-kamilla}
    \Psi_{\tfrac{n}{2pn+1}}(\Omega_{1}, \dots, \Omega_{N}) = \underbrace{
      \begin{vmatrix}
        \LLL Y_{Q^{\ast}, Q^{\ast}, -Q^{\ast}}(\Omega_{1})J_{1} & \dots & \LLL Y_{Q^{\ast}, Q^{\ast}, -Q^{\ast}}(\Omega_{N})J_{N} \\
        \vdots &  \vdots &  \vdots \\
       \LLL Y_{Q^{\ast}, Q^{\ast}, Q^{\ast}}(\Omega_{1})J_{1} & \dots & \LLL Y_{Q^{\ast}, Q^{\ast}, Q^{\ast}}(\Omega_{N})J_{N} \\
        \LLL Y_{Q^{\ast}, Q^{\ast}+1, -(Q^{\ast}+1)}(\Omega_{1})J_{1} & \dots & \LLL Y_{Q^{\ast}, Q^{\ast}+1, -(Q^{\ast}+1)}(\Omega_{N})J_{N} \\
        \vdots & \vdots & \vdots \\
        \LLL Y_{Q^{\ast}, Q^{\ast}+n-1, (Q^{\ast}+n-1)}(\Omega_{1})J_{1} & \dots & \LLL Y_{Q^{\ast}, Q^{\ast}+n-1, (Q^{\ast}+n-1)}(\Omega_{N})J_{N} \\
    \end{vmatrix}}_{\displaystyle \phi_{n}^{\rm JK}(\Omega_{1}, \dots, \Omega_{N};Q^{\ast}+(N-1))}\phi_{1}^{2p-2}.
  \end{equation}
\end{widetext}
This form ensures the wavefunction remains entirely within the LLL: by Eq.~\eqref{eq:lll-harmonics}, the product of any two LLL wavefunctions yields another LLL wavefunction (albeit at different monopole strength). Remarkably, multiple studies have shown that the JK wavefunction is nearly indistinguishable from Eq.~\eqref{eq:cf-ground-state}, despite being formally a different trial wavefunction class~\cite{Jain97, Jain97B, Jain07}. The determinant $\phi_{n}^{\rm JK}$ is often called the CF Slater determinant, with its elements termed CF orbitals.

While we do not discuss the detailed implementation of JK projection, it is instructive to consider the form of a typical JK-projected wavefunction in planar geometry~\cite{Jain97} (after some rearrangement),
\begin{widetext}
  \begin{equation}\label{eq:cf-ground-state-planar-geometry}
    \Psi(z_{1}, \dots, z_{N}) =
    \underbrace{
      \begin{vmatrix}
        1 & \dots & 1 \\
        z_{1} & \dots & z_{N} \\
        \vdots & \vdots & \vdots \\
        e_{1}\left(\left\{\tfrac{1}{z_{1}-z_{j}}\right\}_{j\neq 1}\right) & \dots &
        e_{1}\left(\left\{\tfrac{1}{z_{N}-z_{j}}\right\}_{j\neq N}\right) \\
        \vdots & \vdots & \vdots \\
        z_{1}^{m_{a}+n_{a}} e_{n_{a}}\left(\left\{\tfrac{1}{z_{1}-z_{j}}\right\}_{j\neq 1}\right) & \dots &
        z_{N}^{m_{a}+n_{a}} e_{n_{a}}\left(\left\{\tfrac{1}{z_{N}-z_{j}}\right\}_{j\neq N}\right) \\
        \vdots & \vdots & \vdots
      \end{vmatrix}
    }_{\displaystyle \phi_{\mathrm{JK}}(z_{1}, \dots, z_{N})}
    \prod_{i<j=1}^{N}(z_{i}-z_{j})^{2p}
    \exp\left[-\sum_{i=1}^{N}\frac{|z_{i}|^{2}}{4\ell_{B}^{2}}\right],
  \end{equation}
\end{widetext}
here $z_i = x_i + i y_i$ are the complex coordinates corresponding to the $i^{\rm th}$ electron, $e_{k}$ are the elementary symmetric polynomials and $(n_{a}, m_{a})$ are the quantum numbers corresponding to the occupied $\Lambda$ levels.

Structurally, as seen in Eq.~\eqref{eq:cf-ground-state-planar-geometry}, the JK-projected CF wavefunctions (introduced in 1997) bear an interesting resemblance to Psiformer-like Ans\"atze~\cite{Pfau23} (introduced in 2023, almost a quarter century later): in both cases, the total wavefunction is written as a determinant of coordinate-dependent orbitals $\phi_i(z_i;\{z_{j\neq i}\})$ multiplied by a Jastrow-type correlation factor.

Beyond ground states, CF theory also provides an accurate description of low-lying excitations. (Higher excited states such as electron/hole-like excitations~\cite{Peterson05, Gattu23,Pu23A} can be accessed through the method of CF diagonalization~\cite{Mandal02,Balram13}.) The lowest-energy neutral excitations at filling $\nu=n/(2pn+1)$ are CF particle-hole excitations (CF excitons), created by promoting a CF from the $(n-1)^{\rm th}$ $\Lambda$ level to the $n^{\rm th}$ $\Lambda$ level~\cite{Kamilla96B, Jain07}. We construct an excited state with total angular momentum $(L, M)$ by superposing quasihole-quasiparticle configurations:
\begin{equation}\label{eq:cf-exciton}
  \begin{split}
    &\Psi_{\tfrac{n}{2pn+1}}^{L, M} \\
    &= \LLL \sum_{m_{\rm QH}, m_{\rm QP}}\mathcal{C}_{L_{\rm QH},-m_{\rm QH};L_{\rm QP},m_{\rm QP}}^{L, M}\phi_{n}^{m_{\rm QH}, m_{\rm QP}}\phi_{1}^{2p}.
  \end{split}
\end{equation}
Here, $\phi_{n}^{m_{\rm QH}, m_{\rm QP}}$ is the Slater determinant of $N$ electrons filling $n$ LLs with an electron missing in the $L_{z}=m_{\rm QH}$ orbital of the $(n-1)^{\rm th}$ LL (contributing angular momentum
$L_{\rm QH}=Q^{\ast} + n-1$) and an additional electron in the $L_{z}=m_{\rm QP}$ orbital of the $n^{\rm th}$ LL (contributing $L_{\rm QP}=Q^{\ast} + n$). A quasihole corresponds to a missing CF from the $(n-1)^{\rm th}$ $\Lambda$ level and a quasiparticle corresponds to an additional CF in the $n^{\rm th}$ $\Lambda$ level. The coefficients $\mathcal{C}_{L_{\rm QH},-m_{\rm QH};L_{\rm QP},m_{\rm QP}}^{L, M}$ are Clebsch-Gordan coefficients~\cite{Sakurai20}.

The state with total angular momentum $L = L_{\mathrm{QH}} + L_{\mathrm{QP}}$ and $L_z = L$ contains only a single Fock-space configuration $(m_{\mathrm{QH}}, m_{\mathrm{QP}}) = (-L_{\mathrm{QH}}, L_{\mathrm{QP}})$. In the thermodynamic limit, this state corresponds to an infinitely separated quasihole-quasiparticle pair. The transport gap $\Delta$ is defined as the energy difference between this exciton state and the uniform ground state $\Psi_{\tfrac{n}{2pn+1}}$. This gap governs the activated temperature dependence of the longitudinal 
resistance near the $\nu=n/(2pn+1)$ plateau, $R_{xx} \sim \exp[-\Delta/(2k_B T)]$, with the vanishing of $R_{xx}$ at low temperatures serving as a hallmark of an incompressible FQH state (reflecting conduction only through the ballistic edge modes)~\cite{Girvin85}. In Sec.~\ref{sec:transport-gap}, we use this state as the starting point in the CF-Flow framework to evaluate the $\kappa$-dependence of $\Delta$ in the $\nu = 1/3$ phase.

Having established the CF framework for the idealized LLL limit, we now describe how to combine the backflow approach with CF wavefunctions [Eqs.~\eqref{eq:cf-ground-state} and \eqref{eq:cf-exciton}] to obtain variational Ans\"atze for the lowest energy states within a given angular momentum sector.

\section{The backflow Ansatz}\label{sec:backflow}
In this section, we formulate the backflow approach for the spherical geometry, following the formal construction of Ref.~\cite{Pescia24}.

Let us imagine that we would like to obtain the ground state wavefunction $\Psi_{\rm GS}^{L, M}(\Omega_{1}, \dots, \Omega_{N})$ in the angular momentum sector $(L, M)$ [recall that the eigenstates of the many-body Hamiltonian $\mathcal{H}$ from Eq.~\eqref{eq:many-body-H-kappa} are labelled by $(L, M)$]. Formally, if we know that a state $\Psi_{0}^{L, M}(\Omega_{1}, \dots, \Omega_{N})$—say a state of CFs such as the one in Eq.~\eqref{eq:cf-ground-state} or Eq.~\eqref{eq:cf-exciton}—has a non-zero overlap (however small) with $\Psi_{\rm GS}^{L, M}$, we can obtain the latter from the former through imaginary time evolution. Focusing on the Hamiltonian from Eq.~\eqref{eq:many-body-H-kappa} and dropping the $(L, M)$ indices for visual clarity:
\begin{equation}
  \Psi_{\rm GS}(\Omega_{1}, \dots, \Omega_{N}) = \lim_{\tau \rightarrow \infty} e^{-\tau \mathcal{H}}\Psi_{0}(\Omega_{1}, \dots, \Omega_{N})
\end{equation}

In terms of the infinite-time real-space propagator $\mathcal{G}(\{\Omega\}; \{\Omega^{\prime}\})$, which is the infinite-time transition amplitude between a state of $N$ electrons localized at $\Omega_{1}^{\prime}, \dots, \Omega_{N}^{\prime}$ and one at $\Omega_{1}, \dots, \Omega_{N}$,
\begin{equation}\label{eq:real-space-propagator}
  \mathcal{G}(\{\Omega\}; \{\Omega^{\prime}\}) = \bra{\left\{ \Omega \right\}} \lim_{\tau \rightarrow \infty} e^{-\tau \mathcal{H}} \ket{\left\{ \Omega^{\prime} \right\}}
\end{equation}
we can rewrite the above equation as:
\begin{equation}\label{eq:real-space-propagation}
  \begin{split}
    \Psi_{\rm GS}(\left\{ \Omega \right\}) &\propto \int \prod_i d\Omega^{\prime}_{i}\, \mathcal{G}(\{\Omega\}; \{\Omega^{\prime}\}) \Psi_{0}(\{\Omega^{\prime}\}).
  \end{split}
\end{equation}

Formally, the ground state can thus be obtained by evaluating $\mathcal{G}$. An analytic evaluation is of course impossible, but let us see if we can constrain it in some way.

\begin{figure}
  \includegraphics[width=0.8\columnwidth]{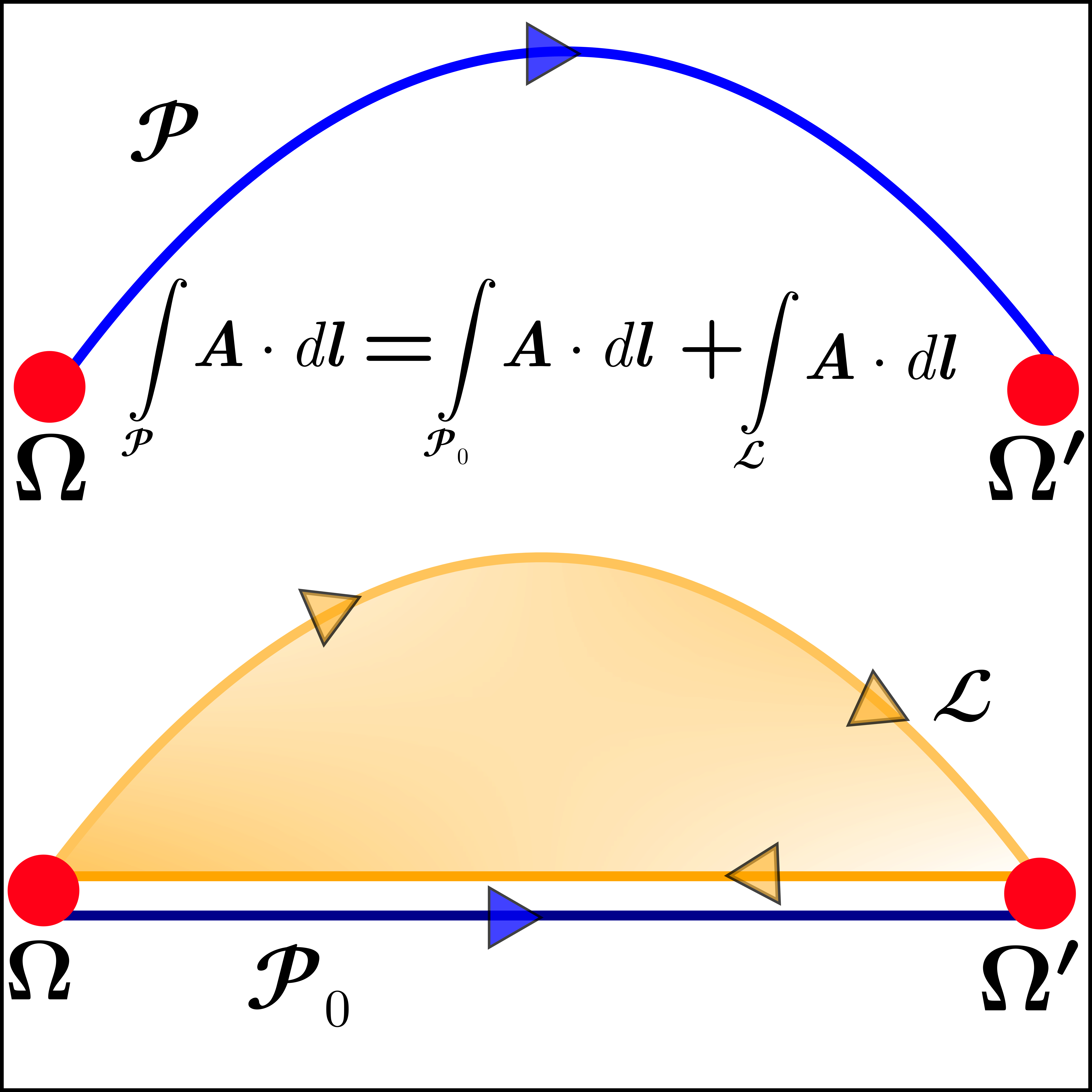}
  \caption{Decomposition of the Aharonov-Bohm phase. The line integral $\int \bm{A}\cdot \bm{dl}$ along the path $\mathcal{P}$ between $\Omega$ and $\Omega^{\prime}$ (top) can always be decomposed into one along a canonical geodesic path $\mathcal{P}_{0}$ and along a closed loop path $\mathcal{L}$ (bottom), where the loop contribution depends only on the enclosed flux—a gauge invariant quantity, which consequently does not depend on the location of $\mathcal{L}$ (since the magnetic field $B$ is uniform everywhere).}
 \label{fig:path-decomposition}
\end{figure}

Let us first note that if the $i^{\rm th}$ electron moves along a path $\mathcal{P}$ between points $\Omega_{i}$ and $\Omega_{i}^{\prime}$, then the wavefunction must acquire a phase: 
\begin{equation}
\exp\left(-\imath \frac{e}{\hbar} \int_{\mathcal{P}} \bm{A}\cdot d\bm{l}\right)
\end{equation}
with $d\bm{l}$ being the line element along $\mathcal{P}$. Imagine, however, that we know only the endpoints of $\mathcal{P}$ [as in Eq.~\eqref{eq:real-space-propagator}]. In this scenario, let us define a canonical path $\mathcal{P}_{0}$ (such as the geodesic connecting $\Omega_{i}$ to $\Omega_{i}^{\prime}$). We can then always decompose the phase acquired by the wavefunction as:
\begin{equation}
  \exp\left(-\imath \frac{e}{\hbar} \int_{\mathcal{P}_{0}} \bm{A}\cdot d\bm{l}\right)\underbrace{\exp\left(-\imath \frac{e}{\hbar}\oint_{\mathcal{L}}\bm{A}\cdot d\bm{l}\right)}_{\mathcal{W}_{\mathcal{L}}}
\end{equation}
Here, $\mathcal{W}_{\mathcal{L}}$ is the phase along some loop (or sum of loops) $\mathcal{L}$—the line integral along any path $\mathcal{P}$ can always be decomposed into the phase acquired along $\mathcal{P}_{0}$ and $\mathcal{W}_{\mathcal{L}}$. The key insight is that $\mathcal{W}_{\mathcal{L}}$ is gauge invariant and equals the flux through $\mathcal{L}$. This flux depends only on the shape of $\mathcal{L}$, not on its location on the sphere as the magnetic field is uniform everywhere. This decomposition is sketched in Fig.~\ref{fig:path-decomposition}.

It therefore follows that we should be able to write $\mathcal{G}$ as:
\begin{equation}
  \begin{split}
    \mathcal{G}(\{\Omega\}; \{\Omega^{\prime}\}) &= \exp\Bigl[\imath \sum_{i}\phi_{\rm AB}(\Omega_{i}^{\prime}, \Omega_{i}) + \mathcal{G}_{\rm re}(\{\Omega\}; \{\Omega^{\prime}\}) \\
    & + \imath \mathcal{G}_{\rm im}(\{\Omega\}; \{\Omega^{\prime}\})\Bigr],
  \end{split}
\end{equation}
where $\phi_{\rm AB}(\Omega_{i}^{\prime}, \Omega_{i})=2Q\arg[\conj{u}_{i}^{\prime}u_{i}+\conj{v}^{\prime}_{i}v_{i}]$ is the Aharonov-Bohm phase accumulated along the geodesic from $\Omega_{i}^{\prime}$ to $\Omega_{i}$ i.e. $\mathcal{P}_{0}$. Here, $\mathcal{G}_{\rm re}$ is an $SO(3)$ invariant function that is even under inversion, while $\mathcal{G}_{\rm im}$ is $SO(3)$ invariant but odd under inversion. The parities under inversion follow from noting that inversion is equivalent to flipping the sign of the magnetic field in the kinetic Hamiltonian (see Eq.~\eqref{eq:kinetic-operator}), i.e., $Q \rightarrow -Q$. Since the wavefunctions must be related to those at the opposite magnetic field via complex conjugation, the parities of $\mathcal{G}_{\rm re}$ and $\mathcal{G}_{\rm im}$ follow. Both functions are equivariant under permutations of electron coordinates.

We are now in a position to arrive at the backflow form. By the mean value theorem, any integral over a compact domain can be replaced by the integrand evaluated at some point within the domain, multiplied by a volume factor. Therefore, there exist ``backflow coordinates'' $\Omega^{\rm BF}_{i} \equiv \Omega_{i}^{\rm BF}(\{\Omega\})$ such that
\begin{equation}
  \begin{split}
    \Psi_{\rm GS}(\left\{ \Omega \right\}) \propto \mathcal{G}(\{\Omega\};\{\Omega^{\rm BF}\})\Psi_{0}(\{\Omega^{\rm BF}\}).
  \end{split}
\end{equation}
The symmetries of the system constrain $\Omega_{i}^{\rm BF}$ to be: (i) permutation equivariant, (ii) $SO(3)$ equivariant, and (iii) odd under inversion—i.e., fully $O(3)$ equivariant. With these constraints, we can replace $\mathcal{G}_{\rm re}(\{\Omega\};\{\Omega^{\rm BF}(\{\Omega\})\})$ and $\mathcal{G}_{\rm im}(\{\Omega\};\{\Omega^{\rm BF}(\{\Omega\})\})$ by simplified functions $\mathcal{J}_{\rm re}(\{\Omega\})$ and $\mathcal{J}_{\rm im}(\{\Omega\})$, which are permutation invariant and $SO(3)$ invariant, with $\mathcal{J}_{\rm re}$ even and $\mathcal{J}_{\rm im}$ odd under inversion.

Before proceeding, we note that $\mathcal{J}_{\rm re}$ faces additional constraints due to the singular Coulomb interaction in Eq.~\eqref{eq:many-body-H-kappa}. Following Kato~\cite{Kato57}, if the $\Psi_{0} \sim d_{ij}^{\mcusp} e^{\imath \mcusp\phi_{ij}}$ at short interparticle distance $d_{ij}=|\Omega_{i}-\Omega_{j}|$—where $\phi_{ij}$ is the azimuthal orientation of the tangent vector connecting $\Omega_{i}$ and $\Omega_{j}$ and $m$ is an odd integer (required by fermionic statistics)—then $\Psi_{\rm GS}$ must satisfy the cusp condition:
\begin{equation}
  \begin{split}
    \lim_{d_{ij} \to 0}\Psi_{\rm GS} \sim e^{\imath \mcusp \phi_{ij}}d_{ij}^{\mcusp}\left(1+\frac{\kappa d_{ij}}{(2\mcusp+1)\lB}\right)
  \end{split}
\end{equation}
Here, we have assumed that the exponent $\mcusp$ as we go from $\Psi_{0}$ to $\Psi_{\rm GS}$ remains unchanged. (This is true in general for any state of CFs except when CFs are restricted entirely to the lowest $\Lambda$ level).

With this in mind, we decompose $\mathcal{J}_{\rm re}$ as $\tilde{\mathcal{J}}_{\rm re} - ({\kappa}/{2\mcusp+1})\sum_{i<j}{\mu^{2}}/{(\mu+d_{ij})}$, where $\tilde{\mathcal{J}}_{\rm re}$ is a function with vanishing first derivative as $d_{ij} \to 0$ and $\mu > 0$ is a gating parameter that preserves $d_{ij}\to 0$ behavior and ensures decay as $d_{ij} \to \infty$. Combining all these elements, the ground state wavefunction takes the form:
\begin{equation}\label{eq:backflow-gs}
\begin{split}
\Psi_{\rm GS}(\{\Omega\}) &= \exp\Biggl[\imath \sum_{i}\phi_{\rm AB}(\Omega_{i}^{\rm BF}, \Omega_{i}) + \tilde{\mathcal{J}}_{\rm re}(\{\Omega\}) \\
&\quad - \frac{\kappa}{2\mcusp+1}\sum_{i<j}\frac{\mu^{2}}{\mu+d_{ij}} + \imath \mathcal{J}_{\rm im}(\{\Omega\})\Biggr] \\
&\times \Psi_{0}(\{\Omega^{\rm BF}\}).
\end{split}
\end{equation}
This formulation provides the foundation for CF-Flow. We choose $\Psi_{0}$ to be a CF wavefunction [Eq.~\eqref{eq:cf-ground-state} or Eq.~\eqref{eq:cf-exciton}], and approximate the backflow coordinates $\Omega_{i}^{\rm BF}$ and Jastrow factors $\tilde{\mathcal{J}}_{\rm re}$, $\mathcal{J}_{\rm im}$ using symmetry-preserving neural networks—the architecture of which we describe in Sec.~\ref{sec:architecture-cf-flow}.

Equation~\eqref{eq:backflow-gs} admits an intuitive physical interpretation closely related to the original backflow picture of Feynman and Cohen~\cite{Feynman56}. Consider a reference Hamiltonian $\mathcal{H}_{0}$ for which $\Psi_{0}$ represents a state of noninteracting CFs. Now imagine adiabatically turning on interactions between CFs—specifically, the additional interactions beyond those already captured by $\mathcal{H}_{0}$. In the semiclassical picture, these interactions induce two effects: First, they cause each CF to experience a displacement from its original position $\Omega_{i}$ to a new effective position $\Omega_{i}^{\rm BF}$, with this displacement depending on the configuration of all other CFs. Second, as the $i^{\rm th}$ CF moves from $\Omega_{i}$ to $\Omega_{i}^{\rm BF}$, it accumulates an Aharonov-Bohm phase $\phi_{\rm AB}(\Omega_{i}^{\rm BF}, \Omega_{i})$ due to the background magnetic field. The Jastrow factors $\tilde{\mathcal{J}}_{\rm re}$ and $\mathcal{J}_{\rm im}$ capture additional correlations induced by the interactions: $\tilde{\mathcal{J}}_{\rm re}$ modifies the wavefunction amplitude (together with the Kato cusp term), while $\mathcal{J}_{\rm im}$ introduces collective phases arising from the many-body dynamics.

We conclude with an important technical point concerning the imaginary terms in Eq.~\eqref{eq:backflow-gs}. As discussed in Sec.~\ref{sec:single-particle}, it is important to ensure that any variational wavefunction is confined entirely to the physical monopole sector. This condition is explicitly enforced by the $\exp[\imath \sum_{i} \phi_{\mathrm{AB}}(\Omega_{i}^{\mathrm{BF}}, \Omega_{i})]$ factor, together with the transformation properties of $\Omega_{i}^{\mathrm{BF}}$ and $\mathcal{J}_{\mathrm{im}}$ under rotations and inversion.

\section{CF-Flow}\label{sec:cf-flow}
We now introduce CF-Flow—a variational framework that combines the backflow approach (Sec.~\ref{sec:backflow}) with CF wavefunctions [such as the ones in Eqs.~\eqref{eq:cf-ground-state} and~\eqref{eq:cf-exciton}] through the use of symmetry-preserving neural networks. We begin with a brief overview of multilayer perceptrons (MLPs) and the DeepSets architecture~\cite{Zaheer17}, which form the building blocks of CF-Flow. We then discuss theorems concerning $O(3)$-equivariant functions that will be useful in constructing our symmetry-preserving Ansatz, before presenting the detailed architecture of CF-Flow.

\subsection{Multilayer perceptrons and DeepSets}\label{sec:mlp-deepsets}
Multilayer perceptrons (MLPs) are universal function approximators~\cite{Carleo19} inspired by neurons and widely used in machine learning. They consist of a sequence of perceptron layers defined as follows.

A perceptron layer maps an input vector $\bm{x} \in \mathbb{R}^{d_{\mathrm{in}}}$ to an output $\bm{y} \in \mathbb{R}^{d_{\mathrm{out}}}$ via
\begin{equation}\label{eq:perceptron-layer}
  \bm{y} = \sigma(\bm{W}\bm{x} + \bm{b}),
\end{equation}
where $\bm{W} \in \mathbb{R}^{d_{\mathrm{out}} \times d_{\mathrm{in}}}$ is the weight matrix, $\bm{b} \in \mathbb{R}^{d_{\mathrm{out}}}$ is the bias vector, and $\sigma$ is a nonlinear activation function applied elementwise. The linear transformation $\bm{W}\bm{x} + \bm{b}$ projects the input into a (typically higher-dimensional) feature space, while the nonlinearity $\sigma$ enables the network to capture interactions between features. 

An MLP is formed by composing multiple such layers,
\begin{equation}\label{eq:multilayer-perceptron}
  \bm{y}^{(I+1)} = \sigma_I(\bm{W}_I \bm{y}^{(I)} + \bm{b}_I),
\end{equation}
with parameters $\{\bm{W}_I, \bm{b}_I\}$ optimized during training. A sufficiently expressive MLP can approximate any continuous function on a compact domain—a property we will exploit in constructing our Ansatz.

However, an MLP's ability to learn a given function depends crucially on the training data. For instance, if the target function is invariant under exchange of its inputs but the data do not exhibit this symmetry, the trained MLP will likewise fail to respect it. In the context of variational Monte Carlo (VMC) optimization~\cite{Lange24}, where only a finite number of samples can be drawn within a finite simulation time, this becomes a significant limitation: unless symmetries are explicitly enforced, the resulting variational Ansatz will not preserve them.

To overcome this, we employ architectures that incorporate the relevant symmetries by construction: the DeepSets framework, which uses MLPs to represent functions over sets—ensuring invariance under permutations of particle indices—and the $O(3)$-equivariant models introduced in Ref.~\cite{Villa21}, which we adapt to our system of indistinguishable electrons (for a broader overview of symmetry-preserving neural-network design, see Refs.~\cite{Bronstein17, Debenedetti25}).

The DeepSets framework~\cite{Zaheer17} allows for the approximation of functions invariant under the exchange of inputs. Specifically, consider a function $f$ that satisfies $f(\bm{x}_{1}, \dots, \bm{x}_{N}) = f(\bm{x}_{\pi(1)}, \dots, \bm{x}_{\pi(N)})$ for any permutation $\pi$ of its inputs. Then $f$ can always be written as
\begin{equation}\label{eq:DeepSets}
f(\bm{x}_{1}, \dots, \bm{x}_{N}) = g\left(\sum_{i} \bm{h}(\bm{x}_{i})\right),
\end{equation}
where $\bm{h}$ and $g$ are suitable functions, and the sum runs over the input index $i$ (not over the components of $\bm{h}$). In the DeepSets framework, both $\bm{h}$ and $g$ are represented using MLPs (see Eq.~\eqref{eq:multilayer-perceptron}). As long as these MLPs are sufficiently wide, this construction can approximate any function $f$ that is invariant under permutations of its inputs~\cite{Zaheer17, Wagstaff21}. We will make use of this property below.

\subsection{\texorpdfstring{$O(3)$-equivariant functions}{O(3)-equivariant functions}}\label{sec:o3-equivariant-functions}

CF-Flow provides a machine-learning approximation to the backflow representation of the ground state $\Psi_{\rm GS}$ in Eq.~\eqref{eq:backflow-gs}. Because evaluating $\tilde{\mathcal{J}}_{\rm re}$, $\mathcal{J}_{\rm im}$, $\mu$, and $\Omega^{\rm BF}$ exactly is infeasible, we approximate each of them using neural networks whose parameters are optimized via variational Monte Carlo (VMC) to minimize the energy. These networks constitute the learnable components of our Ansatz.

The gating parameter $\mu$ is represented by an MLP that maps the physical inputs $\kappa$, $N$, and $Q$ to a positive scalar (see below for details). For the remaining components—$\tilde{\mathcal{J}}_{\rm re}$, $\mathcal{J}_{\rm im}$, and $\Omega^{\rm BF}$—we design architectures that explicitly enforce the symmetry constraints discussed in Sec.~\ref{sec:backflow}.

To this end, we invoke the theory of invariants of Lie groups~\cite{Weyl97}, following Ref.~\cite{Villa21} and adapting it to the case of indistinguishable electrons. The key results~\cite{Weyl97, Villa21} can be summarized as follows:

\textbf{Theorem 1.} If $\bm{v}_{1}, \dots, \bm{v}_{N} \in \mathbb{R}^{3}$, any $SO(3)$-invariant scalar that is even under inversion (i.e., $O(3)$ invariant) can be expressed as a function of dot products $\bm{v}_{i} \cdot \bm{v}_{j}$.

\textbf{Theorem 2.} If $\bm{v}_{1}, \dots, \bm{v}_{N} \in \mathbb{R}^{3}$, any $SO(3)$-invariant scalar that is odd under inversion can be written as a sum over scalar triple products $\chi_{ijk} = \bm{v}_{i} \cdot (\bm{v}_{j} \times \bm{v}_{k})$, multiplied by $O(3)$-invariant functions of dot products $\bm{v}_{a} \cdot \bm{v}_{b}$.

\textbf{Theorem 3.} If $\bm{v}_{1}, \dots, \bm{v}_{N} \in \mathbb{R}^{3}$, any $O(3)$-equivariant vector-valued function can be expressed as a sum over $\bm{v}_{i}$ with $O(3)$-invariant coefficients built from functions of the dot products $\bm{v}_{a} \cdot \bm{v}_{b}$.

Combined with the DeepSets framework~\cite{Zaheer17}—which can approximate any permutation-invariant function using sufficiently expressive MLPs—these theorems provide the foundation for constructing symmetry-preserving networks for each CF-Flow module: $\tilde{\mathcal{J}}_{\rm re}$ (amplitude Jastrow), $\mathcal{J}_{\rm im}$ (phase Jastrow), and $\Omega^{\rm BF}$ (backflow).

\subsection{Architecture of CF-Flow}\label{sec:architecture-cf-flow}
In designing CF-Flow, we embed the physical parameters $\kappa$, $N$, and $Q$ into every learned component, following the ``foundational'' approach introduced in Ref.~\cite{Rende25, Viteritti25}. This enables a single variational Ansatz for the ground state—within the same angular momentum sector as the reference CF wavefunction—to describe systems across a continuous range of LL mixing strengths (we have focused on $\kappa \in (1, 20)$) and system sizes $N$. Embedding $N$ and $Q$ explicitly also opens the possibility of ``transfer learning'': training at smaller system sizes, where optimization is computationally cheaper, and using the resulting model as initialization for larger systems. Throughout this section, we denote by $\bm{r}_{i}$ the Cartesian representation of $\Omega_{i}$, i.e., the coordinates of the $i^{\mathrm{th}}$ electron on the sphere, and by $\bm{R}_{i}$ the Cartesian representation of $\Omega^{\mathrm{BF}}_{i}$—the corresponding backflow coordinate.

\subsubsection{The gating parameter}
As mentioned previously, we approximate $\mu$ using an MLP acting on the inputs $(\kappa, N, Q)$. To enforce positivity, the output layer employs the softplus activation function,
\begin{equation}\label{eq:softplus}
  \text{softplus}(x) = \ln (1+e^{x}),
\end{equation}
so that
\begin{equation}\label{eq:gating-mlp}
  \mu = \text{softplus}\left(\text{MLP}([\kappa, N, Q])\right).
\end{equation}

\subsubsection{Amplitude Jastrow}
The amplitude term $\tilde{\mathcal{J}}_{\mathrm{re}}$ is an $O(3)$-invariant function of $\bm{r}_{1}, \dots, \bm{r}_{N}$ and can therefore be expressed solely in terms of the pairwise distances $d_{ij} = |\bm{r}_{i} - \bm{r}_{j}|$, which are related to the dot products $\bm{r}_{i} \cdot \bm{r}_{j}$ through an invertible mapping. To satisfy the cusp constraint $\partial_{d_{ij}}\tilde{\mathcal{J}}_{\mathrm{re}}|_{d_{ij}=0}=0$, we write
\begin{equation}\label{eq:jastrow-real-model}
  \tilde{\mathcal{J}}_{\mathrm{re}}(\{\bm{r}\}) = \sum_{i<j} f_{ij}(\{d_{ab}\})\, d_{ij}^{2},
\end{equation}
which is always possible for analytic $\tilde{\mathcal{J}}_{\mathrm{re}}$. Furthermore, permutation invariance requires $f_{ij}$ to be symmetric under permutations of $\{\bm{r}_{a} : a \neq i,j\}$, which we enforce via the DeepSets architecture (see Fig.~\ref{fig:model-flowchart}).

\subsubsection{Phase Jastrow}
The phase term $\mathcal{J}_{\mathrm{im}}$ is $SO(3)$-invariant and odd under inversion, and can thus be expressed as a sum over scalar triple products $\chi_{ijk} = \bm{r}_{i} \cdot (\bm{r}_{j} \times \bm{r}_{k})$ with coefficient functions $f_{ijk}(\{d_{ab}\})$:
\begin{equation}\label{eq:jastrow-phase-model}
  \mathcal{J}_{\mathrm{im}}(\{\bm{r}\}) = \sum_{i<j<k} f_{ijk}(\{d_{ab}\})\, \chi_{ijk}.
\end{equation}
Since $\chi_{ijk}$ is totally antisymmetric in $i,j,k$, permutation invariance requires $f_{ijk}$ to also be totally antisymmetric. We construct such an $f_{ijk}$ as a determinant over three $O(3)$-invariant three-dimensional vectors $\bm{f}_{a}^{ijk}$ ($a=i,j,k$), as shown in Fig.~\ref{fig:model-flowchart}.

\subsubsection{Backflow coordinates}
The backflow coordinates $\bm{R}_{i}(\{\bm{r}\})$ are $O(3)$-equivariant. To construct them, we express them as a sum over $\bm{r}_{j}$ with coefficient functions $f_{ij}(\{d_{ab}\})$ that are symmetric under permutations of $\{\bm{r}_{a} : a \neq i,j\}$:
\begin{equation}\label{eq:backflow-sum}
  \bm{R}_{i} = \sum_{j} f_{ij}(\{d_{ab}\})\, \bm{r}_{j}.
\end{equation}
To ensure $\bm{R}_{i}$ lies on the sphere, we first construct a tangent vector
\begin{equation}\label{eq:tangent-vector}
  \bm{v}_{i} = \sum_{j} f_{ij}\, \bm{t}_{ij}, \quad 
  \bm{t}_{ij} = \bm{r}_{j} - (\bm{r}_{j} \cdot \bm{r}_{i})\, \bm{r}_{i},
\end{equation}
where $\bm{t}_{ij}$ projects $\bm{r}_{j}$ onto the tangent plane at $\bm{r}_{i}$, and then apply the exponential map on the sphere~\cite{Perdigao92}:
\begin{equation}\label{eq:expmap}
  \bm{R}_{i} = \exp_{\bm{r}_{i}}(\bm{v}_{i})
  = \cos(v_{i})\, \bm{r}_{i} + \sin(v_{i})\, \hat{\bm{v}}_{i}.
\end{equation}
The coefficients $f_{ij}$ are constructed as before using DeepSets.

\begin{figure*}
  \includegraphics[width=\linewidth]{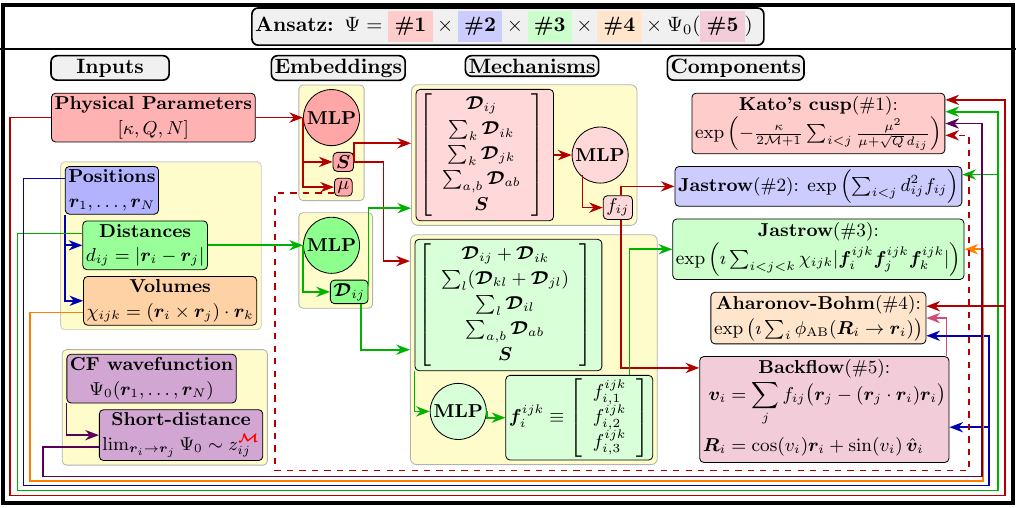}
\caption{Schematic illustration of the CF-Flow architecture. The Ansatz (top) consists of a CF wavefunction $\Psi_{0}$ evaluated at backflow coordinates $\bm{R}_{i}$ and multiplied by four physically motivated factors. The model takes as input the physical parameters $\kappa$, $Q$, and $N$ (Landau-level mixing strength, monopole strength, and particle number) and constructs geometric features from electron coordinates $\bm{r}_{1}, \dots, \bm{r}_{N}$, pairwise distances $d_{ij}$ and signed volumes $\chi_{ijk}$. Using the DeepSets architecture~\cite{Zaheer17}, multilayer perceptrons (MLPs) learn $O(3)$-invariant scalars parameterizing each component: (i) ``two-body'' scalars $f_{ij}$ (many-body functions with pair indices) for an amplitude-modulating Jastrow factor (\#2) and backflow coordinates (\#5), and (ii) ``three-body'' scalars $\bm{f}_{a}^{ijk}$ ($a=i,j,k$; many-body functions with triplet indices) for a phase-modulating Jastrow factor (\#3). The architecture also includes physically enforced terms: Kato's cusp (\#1) with short-distance coefficient set by $\mathcal{M}$ (the relative angular momentum at small $d_{ij}$ in $\Psi_{0}$) and $\kappa$, gated at large distances; and the Aharonov-Bohm phase (\#4) enforcing our gauge choice (in spinor representation, $\phi_{\rm AB}(U, V;u,v)=2Q\arg[\bar{U}u+\bar{V}v]$). By construction, the combined Jastrow term is $SO(3)$ invariant and transforms to its complex conjugate under inversion, while the full wavefunction is manifestly permutation equivariant.}
  \label{fig:model-flowchart}
\end{figure*}

\medskip
Figure~\ref{fig:model-flowchart} summarizes the complete CF-Flow architecture. We now turn to its practical implementation and the results obtained from VMC.

\section{Results}\label{sec:results}
In this section, we first describe the training protocol and hyperparameter choices used to implement CF-Flow. We then present the ground-state energies at $\nu = 1/3$ and $\nu = 2/5$ in the $L=0$ sector (i.e. states with uniform density) as functions of the Landau-level mixing strength $\kappa$ and compare them with earlier results. Finally, we evaluate the transport gap at $\nu = 1/3$ as a function of $\kappa$.

\subsection{Training and hyperparameters}\label{sec:training-hyperparameters}
We optimize the variational parameters $\left\{ \lambda_{\alpha} \right\}$ appearing in the learnable components of CF-Flow using the stochastic reconfiguration (SR) method~\cite{Sorella98, Sorella05}. Following the foundational model approach of Refs.~\cite{Rende25, Viteritti25}, we train a single variational wavefunction that accepts the LL mixing strength $\kappa$ as an input parameter, thereby serving as an Ansatz for the entire family of Hamiltonians $\mathcal{H}(\kappa)$ defined in Eq.~\eqref{eq:many-body-H-kappa}.

Within this framework, parameter updates are determined by ensemble-averaged quantities: the quantum geometric tensor (QGT) $\mathcal{S}_{\alpha \beta}$ and the variational force $\mathcal{F}_{\alpha}$, defined as~\cite{Rende25}
\begin{equation}\label{eq:ensemble-defs}
\begin{split}
\mathcal{F}_{\alpha} &= \frac{1}{\kappa_{\mathrm{max}}-\kappa_{\mathrm{min}}}
\int_{\kappa_{\mathrm{min}}}^{\kappa_{\mathrm{max}}} d\kappa\, F_{\alpha}(\kappa), \\
\mathcal{S}_{\alpha \beta} &= \frac{1}{\kappa_{\mathrm{max}}-\kappa_{\mathrm{min}}}
\int_{\kappa_{\mathrm{min}}}^{\kappa_{\mathrm{max}}} d\kappa\, S_{\alpha \beta}(\kappa),
\end{split}
\end{equation}
where we take $\kappa_{\mathrm{min}} = 1$ and $\kappa_{\mathrm{max}} = 20$.
At each $\kappa$, the force $\mathcal{F}_{\alpha}$ and QGT $\mathcal{S}_{\alpha \beta}$ are computed as~\cite{Rende24}
\begin{equation}\label{eq:force-qgt}
\begin{split}
F_{\alpha} &= -2\,\Re\left[\langle (\mathcal{H}-\langle \mathcal{H}\rangle)
(\mathcal{O}_{\alpha}-\langle \mathcal{O}_{\alpha}\rangle)\rangle\right],\\
S_{\alpha \beta} &= \Re\left[\langle
(\mathcal{O}_{\alpha}-\langle \mathcal{O}_{\alpha}\rangle)^{*}
(\mathcal{O}_{\beta}-\langle \mathcal{O}_{\beta}\rangle)\rangle\right],\\
\mathcal{O}_\alpha &= \partial_{\lambda_{\alpha}}\ln\psi(\mathbf{r}_1,\dots,\mathbf{r}_N;\{\lambda_{\gamma}\}),
\end{split}
\end{equation}
where expectation values are taken with respect to the CF-Flow wavefunction $\psi(\mathbf{r}_1,\dots,\mathbf{r}_N;\{\lambda_{\gamma}\})$.

We evaluate the expectation values through importance sampling with respect to $|\psi|^2$. Specifically, we sample electron configurations using the Metropolis-adjusted Langevin algorithm (MALA), adapted to spherical geometry~\cite{Girolami11}. Compared with the standard random-walk Metropolis-Hastings scheme, MALA achieves equivalent statistical accuracy with significantly fewer samples. At each training iteration (epoch), we draw four $\kappa$ values uniformly from $(1,20)$ to construct the ensemble averages in Eq.~\eqref{eq:ensemble-defs}. For each $\kappa$, we run eight independent Markov chains, generating 400 samples per chain after discarding 512 initial samples for equilibration. To reduce memory overhead, we employ the ``mini-SR'' technique of Ref.~\cite{Rende24}.

The variational parameters are updated via
\begin{equation}
\lambda_{\alpha}(t{+}1)=\lambda_{\alpha}(t)
+\eta(t)\,[\mathcal{S}_{\alpha \beta}(t)+\epsilon(t)\,\delta_{\alpha \beta}]^{-1}\mathcal{F}_{\beta}(t)
\end{equation}
where $\eta(t)$ is the learning rate and $\epsilon(t)$ is a regularization parameter. Both are decayed during training: $\eta(t)$ is halved whenever the (ensemble-averaged) energy plateaus (i.e., ceases to decrease), and optimization terminates after $250$ epochs without improvement.

\begin{figure}
  \includegraphics[width=\linewidth]{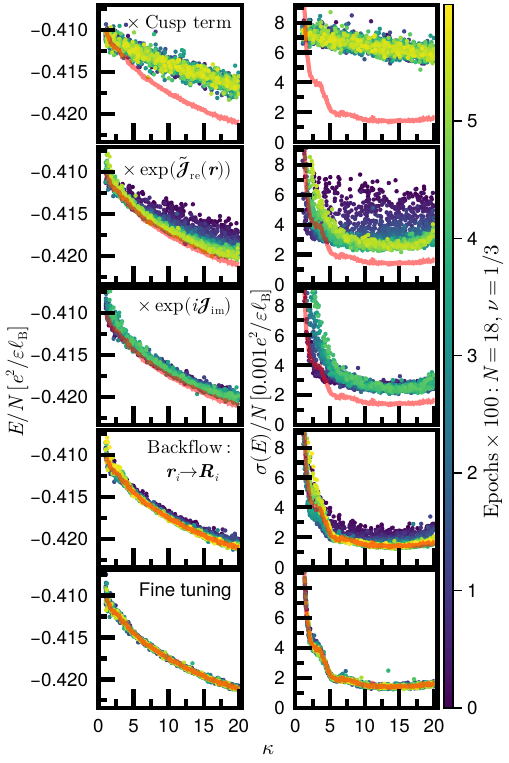}
  \caption{Per-particle energy $E/N$ (left) and local-energy standard deviation $\sigma(E)/N$ (right) as functions of the Landau-level mixing strength $\kappa$ during optimization of CF-Flow for $N=18$ electrons at $\nu=1/3$. Color indicates the training epoch. CF-Flow begins from Laughlin's wavefunction—the uniform ($L=M=0$) state of CFs filling the $n=0$ $\Lambda$ level [see Eq.~\eqref{eq:cf-ground-state}]—and progressively learns backflow corrections through neural networks (see Fig.~\ref{fig:model-flowchart}). Training proceeds in five stages: (i) enabling Kato's cusp term, (ii) adding the amplitude Jastrow factor $\exp[\tilde{\mathcal{J}}_{\rm re}(\mathbf{r})]$, (iii) adding the phase Jastrow factor $\exp[i\mathcal{J}_{\rm im}(\mathbf{r})]$, (iv) enabling backflow, and (v) fine-tuning with a reduced learning rate. The red curve marks the best model (lowest mean energy across $\kappa$ over all training runs). The total energy $E$ includes contributions from the uniform background charge and is corrected for the ``density shift'' inherent to spherical geometry~\cite{Morf86}.}\label{fig:training-curve}
\end{figure}

To accelerate convergence, we employ a staged training protocol in which learnable components are activated sequentially: (i) Kato's cusp term, (ii) amplitude Jastrow, (iii) phase Jastrow, (iv) backflow coordinates, and (v) fine-tuning with reduced learning rate. Each stage initializes from the best parameters of the previous one (determined by the lowest average energy across $\kappa$ during training). Figure~\ref{fig:training-curve} shows a representative training trajectory. The order reflects the increasing complexity of the modules; we have verified that alternative orderings yield equivalent final results.

This staged approach reveals the relative importance of each component to the final energy. The gated Kato cusp term—which decays at large inter-particle separations $d_{ij}$ and connects smoothly to LLL wavefunctions in the $\kappa \to 0$ limit—provides modest improvement, with larger relative gains at smaller $N$. We set $\mathcal{M}=3$ for $\nu=1/3$ (Laughlin) and $\mathcal{M}=1$ for all other states (see Appendix~\ref{appendix:short-distance-constraint}). The amplitude Jastrow and backflow coordinates yield the largest energy gains: the amplitude Jastrow alone should, in principle, reproduce fp-DMC results (which constrains only the phase), while backflow also enables optimization of the phase of the wavefunction~\cite{Luo19}. In contrast, the phase Jastrow term provides negligible improvement. This is likely due to statistical noise in estimating the components of $\bm{S}$ and $\bm{F}$ associated with $\mathcal{J}_{\rm im}$, since the sampling distribution $|\psi|^2$ is insensitive to the phase.

We do not believe the energies and local-energy deviations reported here represent fundamental limits of CF-Flow. Further improvements could be achieved by reducing stochastic noise through longer sampling chains, though we did not pursue this further due to computational cost.

This staged framework also guides the choice of hyperparameters for the individual learnable modules of CF-Flow. We employ a small MLP for $\mu$, while the backflow and amplitude-Jastrow components use larger MLPs with one or two perceptron layers of width $48$-$96$. The phase-Jastrow network is kept roughly six times smaller, as its contribution to the total energy is minor (we have checked this explicitly by considering larger networks). In total, the model contains about $5\times10^{4}$ trainable parameters. Networks with $2$-$4$ times fewer parameters yield statistically similar results, but we adopt the larger configuration to ensure sufficient representational capacity across system sizes and filling factors.

\subsection{Ground state energies}
\label{sec:ground-state-energies}

\begin{figure}
    \includegraphics[width=\linewidth]{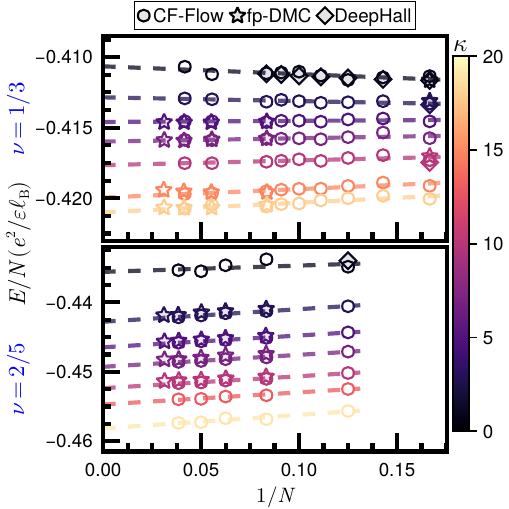}
    \includegraphics[width=\linewidth]{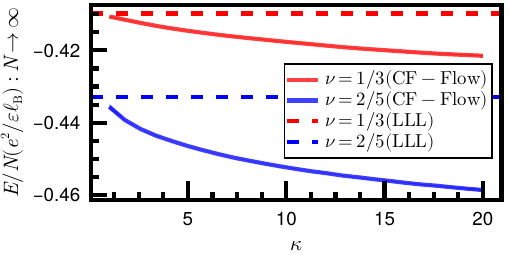}
  \caption{Per-particle energy $E/N$ for uniform (i.e., FQH ground) states at fillings $\nu = 1/3$ and $\nu = 2/5$ as functions of $1/N$ and the Landau-level mixing strength $\kappa$. Top: CF-Flow results compared with DeepHall~\cite{Qian25}—a recent Psiformer-based neural quantum state approach—and fixed-phase diffusion Monte Carlo (fp-DMC)~\cite{Zhao18,Qian25}, which constrains the wavefunction phase to match Eq.~\eqref{eq:cf-ground-state} and provides the ground-state energy within this phase sector. Dashed lines denote linear fits in $1/N$ at fixed $\kappa$. Bottom: Thermodynamic-limit energies $E/N$ ($N \to \infty$) from CF-Flow obtained via linear $1/N$ extrapolation, compared with lowest-Landau-level ($\kappa = 0$) values from CF theory~\cite{Jain07}. Error bars are smaller than the marker size.}
  \label{fig:per-particle-comparisons}
\end{figure}

Figure~\ref{fig:per-particle-comparisons} shows the per-particle energies $E/N$ obtained from CF-Flow at $\nu = 1/3$ and $\nu = 2/5$ starting from the uniform wavefunctions of Eq.~\eqref{eq:cf-ground-state}, i.e., the $\nu^{\ast}=1$ and $\nu^{\ast}=2$ IQH states of CFs. The results correspond to the parameter set yielding the lowest average energy across all sampled $\kappa$ values. In all cases, the average (over $\kappa$) per-particle local-energy standard deviation lies in the range $0.002$--$0.004\,e^{2}/\varepsilon\ell_{B}$ (see, e.g., Fig.~\ref{fig:training-curve}). The small local-energy standard deviation indicates that CF-Flow produces states very close to the true ground state. In the limit $\kappa \to 0$, this deviation grows because even a small component of the wavefunction outside the LLL leads to a large energy change.

For comparison with fp-DMC and DeepHall~\cite{Qian25}, which are evaluated at fixed $\kappa$, we interpolate the CF-Flow energies over $\kappa \in [1,20]$ using Gaussian process regression~\cite{Rasmussen05}, which naturally incorporates Monte Carlo uncertainties and can capture mild non-smooth behavior. As shown, CF-Flow yields per-particle energies comparable to both DeepHall and fp-DMC at $\nu=1/3$, and slightly lower (better) energies at $\nu=2/5$.

\subsection{Transport gap}\label{sec:transport-gap}
\begin{figure}
  \includegraphics[width=\columnwidth]{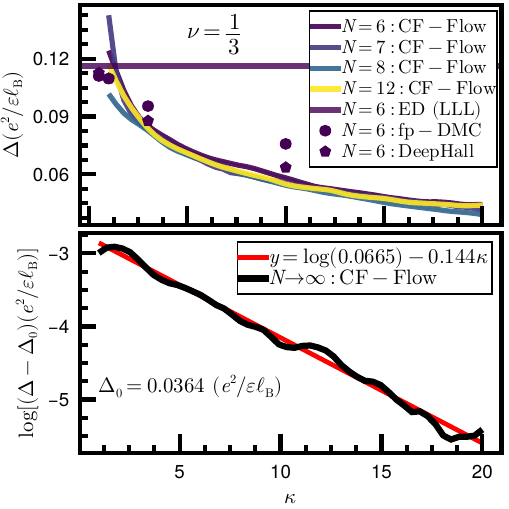}
  \caption{\textbf{Top.} Transport gap $\Delta$ at $\nu=1/3$ as a function of $\kappa$ for various system sizes $N$. The symmetry-preserving architecture of CF-Flow enables construction of variational Ans\"atze for ground or lowest energy states within each angular momentum sector. We compute $\Delta$ as the energy difference between the uniform $L=0$ ground state and the $L=N$ CF exciton state (corresponding to a maximally separated quasihole-quasiparticle pair in the LLL limit, with interactions between separated particles treated as point charges). CF-Flow results are compared with DeepHall~\cite{Qian25}, fixed-phase diffusion Monte Carlo (fp-DMC)~\cite{Zhao18}, and exact diagonalization (ED) in the LLL. \textbf{Bottom.} Thermodynamic-limit gap $\Delta_{\infty}$ obtained via $1/N$ extrapolation (black symbols). The $\kappa$ dependence is well described by an exponential fit (red curve), with $\Delta$ decaying to a finite value ($\Delta_{0}$) in the $\kappa \rightarrow \infty$ limit. (See Eq.~\eqref{eq:transport-gap-kappa}.)}
  \label{fig:transport-gap-1-3}
\end{figure}

To evaluate the transport gap $\Delta$, we initialize CF-Flow from the largest-$L$ CF exciton wavefunction [Eq.~\eqref{eq:cf-exciton}], corresponding to $L = N$ for a system of $N$ electrons at $\nu = 1/3$. This state represents a maximally separated quasihole-quasiparticle pair. The symmetry-preserving design of CF-Flow enables such calculations: the Ansatz is, by construction, restricted to a single total-angular-momentum sector. To reduce computational cost, we focus on $\nu = 1/3$. The results are shown in Fig.~\ref{fig:transport-gap-1-3}.

CF-Flow yields transport gaps comparable to DeepHall~\cite{Qian25} at small $\kappa$, but substantially lower (and therefore more accurate) values at large $\kappa$. The same trend is observed relative to fp-DMC~\cite{Qian25}. Since the ground-state energies from CF-Flow, DeepHall, and fp-DMC remain nearly identical across all $\kappa$, the reduced transport gaps indicate that CF-Flow provides a more accurate description of the excited-state energies than either DeepHall or fp-DMC.

Finally, extrapolating to the thermodynamic limit (Fig.~\ref{fig:transport-gap-1-3}, bottom panel), we find that $\Delta_{\infty}(\kappa)$ is well described by an exponential form,
\begin{equation}\label{eq:transport-gap-kappa}
\Delta_{\infty}(\kappa) = [0.0364 + 0.0665e^{-0.144\kappa}]\frac{e^{2}}{\varepsilon \lB},
\end{equation}
which decays to a finite value $\Delta_{0}$ as $\kappa \to \infty$. This behavior indicates that the system at $\nu = 1/3$ remains incompressible even at large $\kappa$, though the nature of the state may change. The exponential decay of the gap with increasing LL mixing, combined with its convergence to a nonzero value, suggests a first-order transition out of the FQH phase.

\section{Discussion and conclusion}\label{sec:discussion}

Developing a nonperturbative and scalable method to study LL mixing has been a long-standing challenge in the theory of the FQHE. The LLL limit is central to much of our current understanding—most notably through CF theory—but real experimental systems typically operate at $\kappa \sim 0.50-10$~\cite{Sodemann13}. In this nonperturbative regime, LL mixing can drive phase transitions and substantially modify quantitative properties such as the transport gap.

We have introduced CF-Flow, a variational framework that combines CF wavefunctions with backflow transformations implemented through symmetry-preserving neural networks. CF-Flow naturally allows us to address LL-mixing physics—and more generally, any regime where interactions between CFs become significant. By starting from CF wavefunctions, which already encode strong electron correlations, and learning backflow corrections through neural networks, CF-Flow achieves accuracy comparable to or better than recent NQS approaches~\cite{Qian25} while using order of magnitude fewer parameters ($5 \times 10^4$ vs $10^6$) and converging faster ($2 \times 10^{3}$ epochs vs $2 \times 10^{5}$): each calculation for a given system size completes in under two days on a single NVIDIA A100 GPU. This efficiency enables exploration over a substantially larger parameter space in system size, filling factor, and $\kappa$. The symmetry-preserving architecture also gives direct access to excited states by initializing from the appropriate CF wavefunction in any chosen angular-momentum sector.

Beyond its computational advantages, CF-Flow yields insight into the structure of many-body ground states at Jain fillings. At $\nu=1/3$ and $\nu=2/5$, we find that the phase of the IQH wavefunction of CFs [Eq.~\eqref{eq:cf-ground-state}]—which describes the LLL limit ($\kappa\to 0$)—remains remarkably close to that of the true ground state even up to $\kappa \sim 20$. This provides an independent validation of fp-DMC when constrained to this phase and helps explain its quantitative success in predicting experimental properties. However, this phase robustness is not universal: the difference between CF-Flow and fp-DMC transport gaps at $\nu=1/3$ shows that the phase of CF exciton states becomes increasingly sensitive to LL-mixing corrections.

The foundational training strategy of CF-Flow—in which a single variational Ansatz is optimized across continuous ranges of $\kappa$, system size $N$, and monopole charge $Q$—offers both practical and conceptual advantages. It reduces computational cost while maintaining high accuracy across an entire family of Hamiltonians and naturally enables transfer learning across nearby system sizes and, potentially, filling factors. This framework also permits computation of geometric-response quantities such as the fidelity susceptibility with respect to $\kappa$~\cite{Rende24, GU10}, which can identify continuous phase transitions without requiring a predefined order parameter and may give insight into the associated universality class. This capability is particularly valuable in the FQHE: a CF is a highly nonlocal bound state involving all electrons, so order parameters for transitions between distinct CF-driven phases are likely to be similarly nonlocal or difficult to construct. An order-parameter-agnostic approach is therefore especially well suited.

Several directions warrant further exploration. First, we have focused entirely on uniform ($L=0$) states, which represent the ground state within the FQH phase. At sufficiently large $\kappa$, however, the system may transition to a symmetry-breaking state such as a charge-density wave or Wigner crystal. The situation is subtle: such phases may appear as ground states in higher-angular-momentum sectors—if rotational symmetry is spontaneously broken—or they may remain in the $L=0$ sector as rotating crystals that preserve overall rotational invariance. Because CF-Flow respects all symmetries by construction, it can probe both possibilities by comparing energies across different $L$ sectors. Identifying a transition from an FQH liquid to a crystalline phase requires studying spatial correlations. Pair-correlation functions can detect translational order, while distinguishing a rotating crystal from a pinned crystal demands three-body (or higher) correlations that simultaneously fix a lattice site and the orientation of the crystal axes. Although we have not pursued this here, CF-Flow's ability to produce high-quality wavefunctions in arbitrary angular-momentum sectors makes this a natural next step.
Second, real experimental systems contain disorder, which can strongly modify transport properties and excitation gaps. Extending CF-Flow to incorporate disorder—something the foundational approach can accommodate naturally~\cite{Rende25,Viteritti25}—and understanding how disorder and LL mixing together shape observed transport gaps represents an important future direction.

Finally, the recent observation of the fractional quantum anomalous Hall effect (FQAHE)—FQHE in the absence of a magnetic field in Chern bands~\cite{Cai23,Park23,Zeng23,Xu23,Lu24,Bernevig25}—is an exciting area for applying CF-Flow. The prominent FQH states again occur at the Jain fillings $\nu = n/(2n+1)$ and follow a qualitative trend similar to Landau levels, with smaller $n$ producing more robust states (as at these fillings CFs feel a larger effective field). However, it is a priori unclear whether CFs remain weakly interacting in these systems~\cite{Lu23A}. Extending CF-Flow to this setting, and understanding if and how CFs are dressed in Chern bands, would be an exciting future direction.

\section*{Acknowledgements}
M.G. thanks Jainendra K. Jain for many thoughtful discussions throughout the course of this project and for valuable comments on the manuscript. M.G. also thanks Dezhe Jin, Puneeth Nallan Chakravartula, and G. J. Sreejith for insightful discussions and suggestions. M.G. acknowledges support from the National Science Foundation under Grant No.~DMR-2404619, awarded to Jainendra K. Jain. The author of this work recognizes the Penn State Institute for Computational and Data Sciences (RRID:~SCR\_025154) for providing access to computational research infrastructure within the Roar Core Facility (RRID:~SCR\_026424). Plots were made using \texttt{Makie.jl}~\cite{Makie}. All variational Monte Carlo computations in this work were implemented using \texttt{JAX}~\cite{jax18}.

\appendix

\section{Short-distance constraint on CF-Flow}\label{appendix:short-distance-constraint}
In this appendix, we discuss a formal constraint on CF-Flow arising from our choice of activation function and its implications for the short-distance behavior of the wavefunction.

\subsection{Constraint from analyticity}

While the backflow Ansatz in Eq.~\eqref{eq:backflow-gs} is formally universal, our implementation employs smooth activation functions to ensure numerical stability during optimization. The stochastic reconfiguration (SR) method requires second-order spatial derivatives of the wavefunction [see Eq.~\eqref{eq:kinetic-operator}] to evaluate the local kinetic energy. We therefore use the smooth GELU activation function in all MLPs (see Fig.~\ref{fig:model-flowchart}),
\begin{equation}\label{eq:gelu-activation}
\mathrm{GELU}(x) = x\frac{1 + \mathrm{erf}(x/\sqrt{2})}{2},
\end{equation}
which renders the overall Ansatz analytic in the particle coordinates $\mathbf{r}_{i}$.

Because the wavefunction is analytic, CF-Flow cannot modify the order of zeros when two particles coincide. Specifically, it cannot reduce the relative--angular--momentum exponent $\mathcal{M}$ that governs the short-distance behavior $\psi(\mathbf{r}) \sim |\mathbf{r}_i - \mathbf{r}_j|^{\mathcal{M}}$ as $\mathbf{r}_i \to \mathbf{r}_j$. We therefore set $\mathcal{M}$ to match the underlying CF state: $\mathcal{M}=3$ for the $\nu=1/3$ Laughlin state [Eq.~\eqref{eq:cf-ground-state}] and $\mathcal{M}=1$ for all other CF states considered here. (More generally, for two-vortex CF states, as long as at least one CF resides in an $n\ge 1$ $\Lambda$ level after LLL projection, we have $\mathcal{M}=1$.) Higher-vortex ($p\ge 2$) CF states would require different choices, though we have not explored them in this work.

\subsection{Practical implications}

The practical implications of this constraint are not fully understood. Even within the LLL, the Laughlin wavefunction---whose short-distance exponent is fixed at $\mathcal{M}=3$---is known to approximate the Coulomb ground state extremely well: it reproduces the per-particle energy to within $\sim 10^{-4}\, e^{2}/\varepsilon\ell_{B}$~\cite{Laughlin83, Jain07}. At the same time, the exact ground state contains a small admixture of the $\nu=1$ IQH state of CFs and CF excitons, which lowers the short-distance exponent to $\mathcal{M}=1$~\cite{Jain07, Peterson03}. CF-Flow, by construction, retains the exponent of the underlying CF state.

Whether enforcing $\mathcal{M}=3$ at $\nu=1/3$ affects the \emph{reachable} accuracy of CF-Flow is unclear. In practice, we believe the precision is limited by stochastic noise in Monte Carlo sampling and finite-precision effects in gradient-based optimization. Within these limits, any consequence of the fixed exponent would be extremely difficult to resolve.

As shown in Fig.~\ref{fig:per-particle-comparisons}, CF-Flow nevertheless attains the same level of energy accuracy as existing methods. Determining whether altering the short-distance exponent could yield improvements below the current numerical noise floor remains an open question.

\bibliographystyle{unsrtnat}
\bibliography{biblio_fqhe_processed}
\end{document}